\documentclass[aps,prl,10pt,twocolumn,superscriptaddress,nofootinbib,preprintnumbers,showpacs]{revtex4-2}
\usepackage{amsmath,amsfonts, amssymb, amsthm, dsfont}
\usepackage[T1]{fontenc}
\usepackage{yfonts}
\usepackage{bm}
\usepackage{bbm}
\usepackage{mathrsfs}
\usepackage{graphicx}
\usepackage{verbatim}

\usepackage{wasysym}
\usepackage[dvipsnames]{xcolor}
\usepackage{bbold}
\usepackage{braket}
\usepackage{diagbox} 
\usepackage{environ} 
\usepackage{relsize} 
\usepackage[caption=false]{subfig} 
\usepackage{tikz}
\usepackage{ulem}
\usepackage{fontawesome5}
\usepackage{slashed}

\usepackage[bookmarks=true,colorlinks=true,linkcolor=blue, urlcolor=blue, citecolor=blue]{hyperref}
\hypersetup{linktocpage}

\newcommand{\cube}{\text{\scalebox{0.8}{\faDiceD6}}}

\newcommand{\mysec}[1]{\textit{{\textcolor{MidnightBlue}{#1}}}}

\begin{document}

\title{Corner Charge Fluctuations in Higher Dimensions}


\author{Xiao-Chuan Wu}
\affiliation{Department of Physics, Princeton University, Princeton, New Jersey 08544, USA}

\author{Pok Man Tam}
\affiliation{Princeton Center for Theoretical Science, Princeton University, Princeton, New Jersey 08544, USA}

\author{Xuyang Liang}
\affiliation{Guangdong Provincial Key Laboratory of Magnetoelectric Physics and Devices, State Key Laboratory of Optoelectronic Materials and Technologies, Institute of Neutron Science and Technology, School of Physics, Sun Yat-Sen University, Guangzhou, 510275, China}

\author{Zenan Liu}
\affiliation{Department of Physics, School of Science and Research Center for Industries of the Future, Westlake University, Hangzhou 310030,  China}
\affiliation{Institute of Natural Sciences, Westlake Institute for Advanced Study, Hangzhou 310024, China}

\author{Dao-Xin Yao}
\affiliation{Guangdong Provincial Key Laboratory of Magnetoelectric Physics and Devices, State Key Laboratory of Optoelectronic Materials and Technologies, Institute of Neutron Science and Technology, School of Physics, Sun Yat-Sen University, Guangzhou, 510275, China}

\author{Zheng Yan}
\affiliation{Department of Physics, School of Science and Research Center for Industries of the Future, Westlake University, Hangzhou 310030, China}
\affiliation{Institute of Natural Sciences, Westlake Institute for Advanced Study, Hangzhou 310024, China}

\author{Shinsei Ryu}
\affiliation{Department of Physics, Princeton University, Princeton, New Jersey 08544, USA}

\begin{abstract}

Measuring charge fluctuations within a subregion provides a powerful probe of quantum many-body systems. In two spatial dimensions, the shape dependence of the dimensionless corner contribution encodes universal data of quantum critical points and reveals observables of quantum geometry in various quantum phases. Here, we systematically extend this framework to higher dimensions. In three dimensions, we derive the universal angle dependence associated with trihedral corners of a generic parallelepiped and benchmark the predictions against Monte Carlo simulations of lattice models at the O(3) quantum critical point. We further identify a wedge-corner contribution that directly probes the quantum metric, supported by numerical results for a lattice Weyl semimetal model. More generally, we obtain angle functions for polyhedral corners of arbitrary parallelotopes in general dimensions and clarify the scaling of the corner contribution across phases of matter. While insulators and conformal critical points exhibit similar behavior across dimensions, metals display a characteristic even-odd dimensional effect.




\end{abstract}

\maketitle

\mysec{Introduction.} Understanding extended objects that reveal symmetry properties~\cite{gene_sym_review1,gene_sym_review2} and entanglement signatures~\cite{Eisert2010RMP,EECMP2016,Wen2017RMP} has become a central theme in contemporary many-body physics. For systems with a U(1) global symmetry, the disorder operator~\cite{DO_old_1,DO_old_2,DO_fradkin} created by a symmetry twist inside a subregion, together with its small-twist-angle limit given by bipartite fluctuations of the conserved charge, provides an informative probe of the intrinsic and universal data of many-body states. 

Significant progress has been made in two spatial dimensions. When the boundary of the subregion is not smooth but contains a corner, a dimensionless corner contribution generally appears, exhibiting a universal dependence on the corner angle~\cite{estienne2022cornering,Dirac_log,wu2021universal,Wang2021scaling,estienne2022cornering,Wu2025CFS,Wu2025Corner,Tam2024Corner}. In conformal field theories (CFTs)~\cite{Dirac_log,wu2021universal,Wang2021scaling,estienne2022cornering}, as well as in certain quantum-critical metals without conformal symmetry~\cite{Wu2025CFS}, this contribution scales logarithmically with the linear size of the subregion, and its universal coefficient is directly tied to quantum-critical transport properties. These field-theoretical predictions can be readily tested using Monte Carlo simulations of lattice models, and have been benchmarked in the Bose-Hubbard model~\cite{Wang2021scaling} and applied to studies of symmetric mass generation~\cite{Liu2024Disorder} and other unconventional quantum phase transitions~\cite{zhao2022scaling,Jiang2022Many,liu2022measuring,Liu2023Fermion}.

In two-dimensional insulators, the dimensionless corner contribution to charge fluctuations, which does not scale with the subregion size, provides a direct probe of the many-body quantum metric~\cite{Tam2024Corner,Wu2025Corner} defined through adiabatic flux insertion (equivalently, twisted boundary conditions). The quantum metric generally takes non-universal values in a gapped phase, yet is nevertheless universally bounded by the topological invariant given by the Hall conductivity~\cite{SWM2000,Resta2002Rev,Resta2011Rev,Onishi2024Bound2,Wu2025Corner,Onishi2025quantum}.


Beyond two dimensions ($d=2$), our understanding of corner contributions to charge fluctuations remains limited. Only a few subregion geometries in three dimensions, such as spherical, cubic, and conical ones, have been analyzed in CFTs~\cite{Dirac_log,liang2025scaling}, exhibiting similarities to the shape dependence of entanglement entropies~\cite{Solodukhin_2008,CHM_2011,Myers_Singh_2012,klebanov2012shape,Faulkner2016,4D_corner_1,4D_corner_2,4D_corner_3,4D_corner_4,4D_corner_5,EE_singular_2019}. Motivated by the need for analytical results that can guide future numerical studies of lattice models for quantum phases and phase transitions, it is natural to consider lattice-compatible cuts such as a parallelepiped. However, to the best of our knowledge, the angle dependence associated with trihedral corners is currently unavailable in the literature, in neither entanglement entropies (see, e.g., Ref.~\cite{EE_singular_2019}) nor bipartite fluctuations~\cite{Dirac_log,liang2025scaling}. 

Moreover, the connection to quantum geometry does not directly carry over to $d=3$: the trihedral corner contribution remains dimensionless, whereas the many-body quantum metric becomes dimensionful. This mismatch motivates the development of a new protocol for extracting quantum geometry in higher dimensions.



In this Letter, we present a systematic generalization of the existing two-dimensional results to three and higher dimensions. We show that the trihedral corners of a generic parallelepiped admit a universal angle function, which we determine analytically and corroborate by Monte Carlo simulations of lattice models with (3+1)D $\textrm{O}(3)$ quantum criticality. In addition, we demonstrate that the wedge-corner contribution serves as a quantum-geometric observable, supported by numerical calculations on a tight-binding Weyl semimetal. These developments naturally culminate in general formulas applicable to arbitrary spatial dimensions.

\mysec{Preliminaries.} For many-body systems with a U(1) global symmetry in $d$ spatial dimensions, the disorder operator associated with a subregion $\Sigma$ is defined as
\begin{flalign}
\mathcal{U}_{\Sigma}(\chi)=\exp\left(\mathtt{i}\chi\int_{\Sigma}\textrm{d}^{d}\boldsymbol{r}\rho(\boldsymbol{r})\right),
\label{eq:_dis_op}
\end{flalign}
where $\rho$ denotes the U(1) charge density, and $\chi$ is a real-valued parameter. The expectation value $\langle\mathcal{U}_{\Sigma}(\chi)\rangle$ serves as a generating function, and its second cumulant is
\begin{flalign}
\mathcal{F}_{\Sigma}=\lim_{\chi\rightarrow0}(-\mathtt{i}\partial_{\chi})^{2}\log\langle\mathcal{U}_{\Sigma}(\chi)\rangle.
\label{eq:_2nd_cumulant}
\end{flalign}
This quantity, known as the bipartite charge fluctuations, controls the disorder
operator $\langle\mathcal{U}_{\Sigma}(\chi)\rangle\approx1-(\chi^{2}/2)\mathcal{F}_{\Sigma}$ in the small-$\chi$ limit. 

\begin{figure}
    \centering
 \includegraphics[width=1\linewidth]{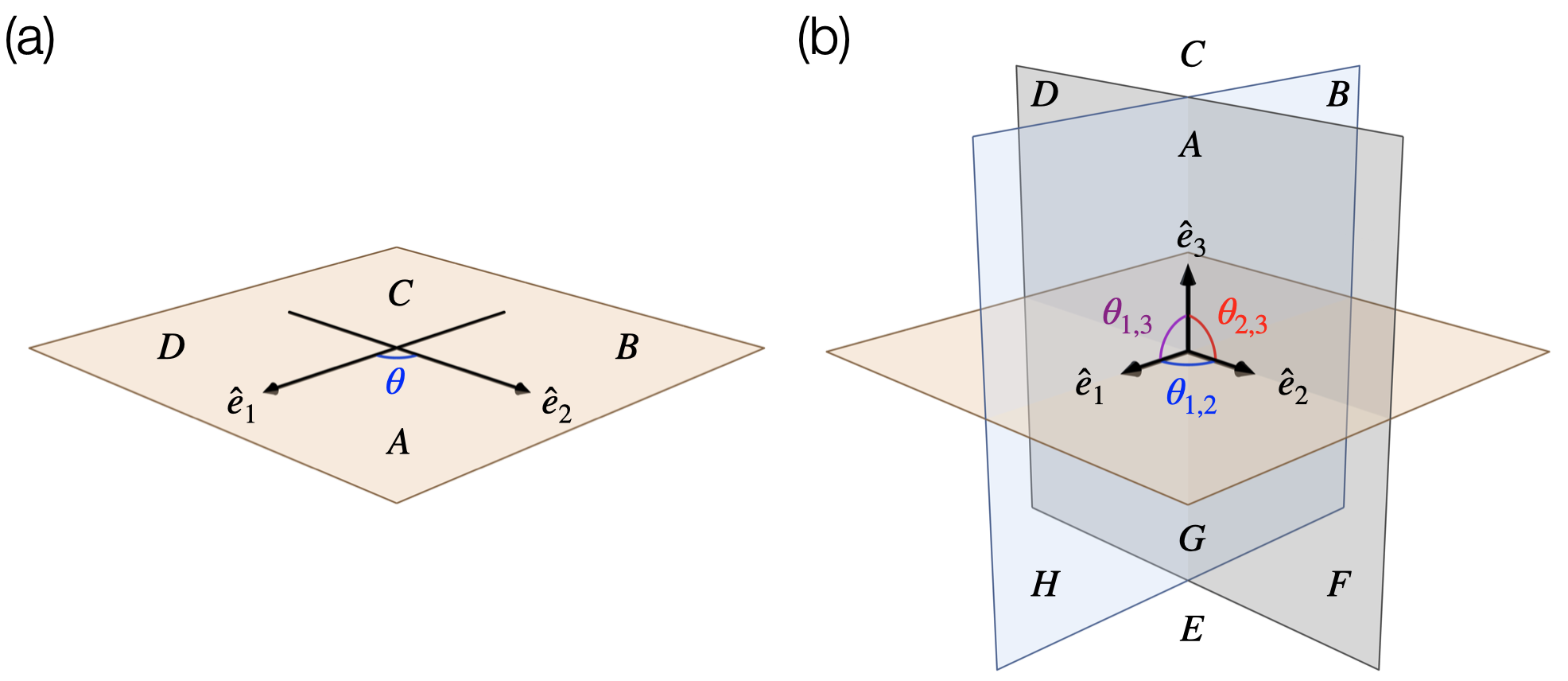}
    \caption{Real-space partitions used to extract corner charge fluctuations in two and three dimensions.}
    \label{fig:_corner}
\end{figure}

Let us briefly review the known results in $d=2$. In both CFTs and insulators, the bipartite fluctuations scale as $\mathcal{F}_{\Sigma}=\#L+\mathcal{C}$, where $L$ is the linear size of $\Sigma$, $\#$ is a non-universal coefficient, and $\mathcal{C}$ is a dimensionless term. Following the subtraction scheme introduced in Ref.~\cite{estienne2022cornering,Tam2024Corner}, the dimensionless contribution associated with an arbitrary parallelogram can be extracted as $\mathcal{C}_{2}=\mathcal{F}_{\textrm{A}}+\mathcal{F}_{\textrm{B}}+\mathcal{F}_{\textrm{C}}+\mathcal{F}_{\textrm{D}}-\mathcal{F}_{\textrm{AB}}-\mathcal{F}_{\textrm{CD}}-\mathcal{F}_{\textrm{BC}}-\mathcal{F}_{\textrm{AD}}+\mathcal{F}_{\textrm{ABCD}}$, where $\textrm{A},\textrm{B},\textrm{C}$ and $\textrm{D}$ denote the four regions in the partition of the plane shown in FIG.~\ref{fig:_corner}. (a). For rotationally invariant systems, one finds~\cite{estienne2022cornering} $\mathcal{C}_{2}=f_{2}(\theta)\varUpsilon_{2}$, where 
\begin{flalign}
f_{2}(\theta)=-(\mathtt{f}(\theta)+\mathtt{f}(\pi-\theta)).
\label{eq:_angle_fun_2d}
\end{flalign}
Here, $\mathtt{f}(\theta)=1+(\pi-\theta)\cot\theta$ is the universal angle function in the literature, and the dimensionless coefficient 
\begin{flalign}
\varUpsilon_{2}=-\int_{a}^{L}\textrm{d}rr^{3}S(r)
\label{eq:_2th_mom}
\end{flalign}
is determined by the (equal-time) static structure factor (SSF) $S(\boldsymbol{r})=\langle\rho(\boldsymbol{r})\rho(0)\rangle_{c}$, where $a$ denotes a short-distance UV cutoff. The quantity $\varUpsilon_{2}$ is finite in insulators and directly related to the many-body quantum metric~\cite{Wu2025Corner,Tam2024Corner}, and it diverges logarithmically at conformal quantum critical points~\cite{estienne2022cornering,Wu2025Corner}.

For a purpose that will become clear later, the universal angle function in Eq.~\eqref{eq:_angle_fun_2d} can be written as $\mathtt{f}(\theta)/2=I_{2}(G^{\{+-\}})=I_{2}(G^{\{-+\}})$ and $\mathtt{f}(\pi-\theta)/2=I_{2}(G^{\{++\}})=I_{2}(G^{\{--\}})$.
Here, $G$ is a Gram matrix defined by 
\begin{flalign}
G_{ij}^{\{\mathbf{s}\}}=s_{i}s_{j}\hat{\boldsymbol{e}}_{i}\cdot\hat{\boldsymbol{e}}_{j}=s_{i}s_{j}\cos(\theta_{i,j})\quad\textrm{with}\quad s_{i}=\pm,
\label{eq:_Gram_matrix}
\end{flalign}
where $\hat{\boldsymbol{e}}_{i}$ are the unit vectors shown in FIG.~\ref{fig:_corner}. (a), and the $\theta_{i,j}$ denotes the angle between $\hat{\boldsymbol{e}}_{i}$ and $\hat{\boldsymbol{e}}_{j}$. The functional $I_{2}(G)$ for a two-dimensional matrix is defined as
\begin{flalign}
I_{2}(G)=\det(G)\int_{0}^{1}\textrm{d}v_{1}\frac{v_{1}v_{2}}{(v^{\mathsf{T}}Gv)^{2}}
\label{eq:_Int_Gram_2d}
\end{flalign}
with $v_{2}=1-v_{1}$.

\mysec{Generalized universal angle function.} We now generalize the analysis to three spatial dimensions.
For an arbitrary parallelepiped $\cube$, sufficiently local charge correlations—as realized in insulators and CFTs—imply the scaling
$\mathcal{F}_{\cube}=(\#_{2})L^{2}+(\#_{1})L+\mathcal{C}_{3}$, 
where $L$ is the linear size of $\cube$ and $\#_{1},\#_{2}$ are non-universal coefficients.
The case of metals, where charge correlations decay more slowly, will be discussed later.
As a generalization of the subtraction scheme in FIG.~\ref{fig:_corner}. (a), the dimensionless contribution $\mathcal{C}_{3}$ can be extracted by
\begin{flalign}
&\mathcal{C}_{3}=\mathcal{F}_{\textrm{A}}+\mathcal{F}_{\textrm{B}}+\mathcal{F}_{\textrm{C}}+\mathcal{F}_{\textrm{D}}+\mathcal{F}_{\textrm{E}}+\mathcal{F}_{\textrm{F}}+\mathcal{F}_{\textrm{G}}+\mathcal{F}_{\textrm{H}}-\mathcal{F}_{\textrm{AB}}\nonumber\\&-\mathcal{F}_{\textrm{AD}}-\mathcal{F}_{\textrm{AE}}-\mathcal{F}_{\textrm{BC}}-\mathcal{F}_{\textrm{CD}}-\mathcal{F}_{\textrm{CG}}-\mathcal{F}_{\textrm{BF}}-\mathcal{F}_{\textrm{EF}}-\mathcal{F}_{\textrm{FG}}\nonumber\\&-\mathcal{F}_{\textrm{DH}}-\mathcal{F}_{\textrm{EH}}-\mathcal{F}_{\textrm{GH}}+\mathcal{F}_{\textrm{ABCD}}+\mathcal{F}_{\textrm{EFGH}}+\mathcal{F}_{\textrm{ABEF}}\nonumber\\&+\mathcal{F}_{\textrm{CDGH}}+\mathcal{F}_{\textrm{ADEH}}+\mathcal{F}_{\textrm{BCFG}}-\mathcal{F}_{\textrm{ABCDEFGH}},
\label{eq:_corner_scheme_3d}
\end{flalign}
where $\textrm{A},\textrm{B},\ldots,\textrm{H}$ label the eight regions depicted in FIG.~\ref{fig:_corner}. (b). In terms of the SSF $S(\boldsymbol{r})$, this becomes $\mathcal{C}_{3}=-2(\Xi_{\textrm{AG}}+\Xi_{\textrm{BH}}+\Xi_{\textrm{CE}}+\Xi_{\textrm{DF}})$, where the pairing $\Xi_{\textrm{AG}}$ between two regions is defined as
\begin{flalign}
\Xi_{\textrm{AG}}=\int_{\textrm{A}}\textrm{d}^{3}\boldsymbol{r}_{1}\int_{\textrm{G}}\textrm{d}^{3}\boldsymbol{r}_{2}S(\boldsymbol{r}_{1}-\boldsymbol{r}_{2}).
\label{eq:_SSF_pair_3d}
\end{flalign}
Since the two regions $\textrm{A}$ and $\textrm{G}$ share only the trihedral vertex, any UV singularity contained in $\Xi_{\textrm{AG}}$ must be attributed to the trihedral corner contribution.

Carrying out the integral in Eq.~\eqref{eq:_SSF_pair_3d} for rotationally invariant systems (see the Supplemental Material (SM)~\cite{SM} for details), we find that $\Xi_{\textrm{AG}}=I_{3}(G^{\{+++\}})\varUpsilon_{3}$. Here, the dimensionless quantity generalizing Eq.~\eqref{eq:_2th_mom} is 
\begin{flalign}
\varUpsilon_{3}=-\int_{a}^{L}\textrm{d}rr^{5}S(r).
\label{eq:_3th_mom}
\end{flalign}
We continue to use the definition Eq.~\eqref{eq:_Gram_matrix} of the Gram matrix in terms of the three unit vectors in FIG.~\ref{fig:_corner}. (b). The three-dimensional generalization of Eq.~\eqref{eq:_Int_Gram_2d} reads
\begin{flalign}
I_{3}(G)=\det(G)\int_{\Delta_{2}}\textrm{d}v_{1}\textrm{d}v_{2}\frac{v_{1}v_{2}v_{3}}{(v^{\mathsf{T}}Gv)^{3}},
\label{eq:_Int_Gram_3d}
\end{flalign}
where $\Delta_{2}$ is the 2-simplex defined by $v_{1}+v_{2}+v_{3}=1$ with $v_{1},v_{2},v_{3}\geq0$. A similar analysis applies to $\Xi_{\textrm{BH}}$, $\Xi_{\textrm{CE}}$, and $\Xi_{\textrm{DF}}$. Combining all contributions in Eq.~\eqref{eq:_corner_scheme_3d} gives $\mathcal{C}_{3}=f_{3}(\theta_{i,j})\varUpsilon_{3}$, where the universal angle dependence for a trihedral corner is
\begin{flalign}
f_{3}(\theta_{1,2},\theta_{1,3},\theta_{2,3})=\sum_{s_{1},s_{2},s_{3}=\pm}I_{3}(G^{\{s_{1},s_{2},s_{3}\}}).
\label{eq:_angle_fun_3d-I}
\end{flalign}
For the special case $\theta_{1,2}=\theta$ and $\theta_{1,3}=\theta_{2,3}=\pi/2$, this expression reduces to
\begin{flalign}
f_{3}(\theta,\pi/2,\pi/2)=1+\left(\frac{\pi}{2}-\theta\right)\cot\theta.
\label{eq:_angle_fun_3d-II}
\end{flalign}

The behavior of the dimensionless quantity in Eq.~\eqref{eq:_3th_mom}, as well as its higher-dimensional generalizations, depends on the underlying phase of matter. As a simple illustration, the SSF scales as $S(r)=-C_{J}/r^{6}$ in CFTs, leading to $\varUpsilon_{3}=C_{J}\log (L/a)$ controlled by the current central charge $C_{J}$ (see the SM~\cite{SM}). In the following section, we test this prediction using quantum Monte Carlo (QMC) simulations of lattice models.



\begin{figure}
    \centering
    \includegraphics[width=0.8\linewidth]{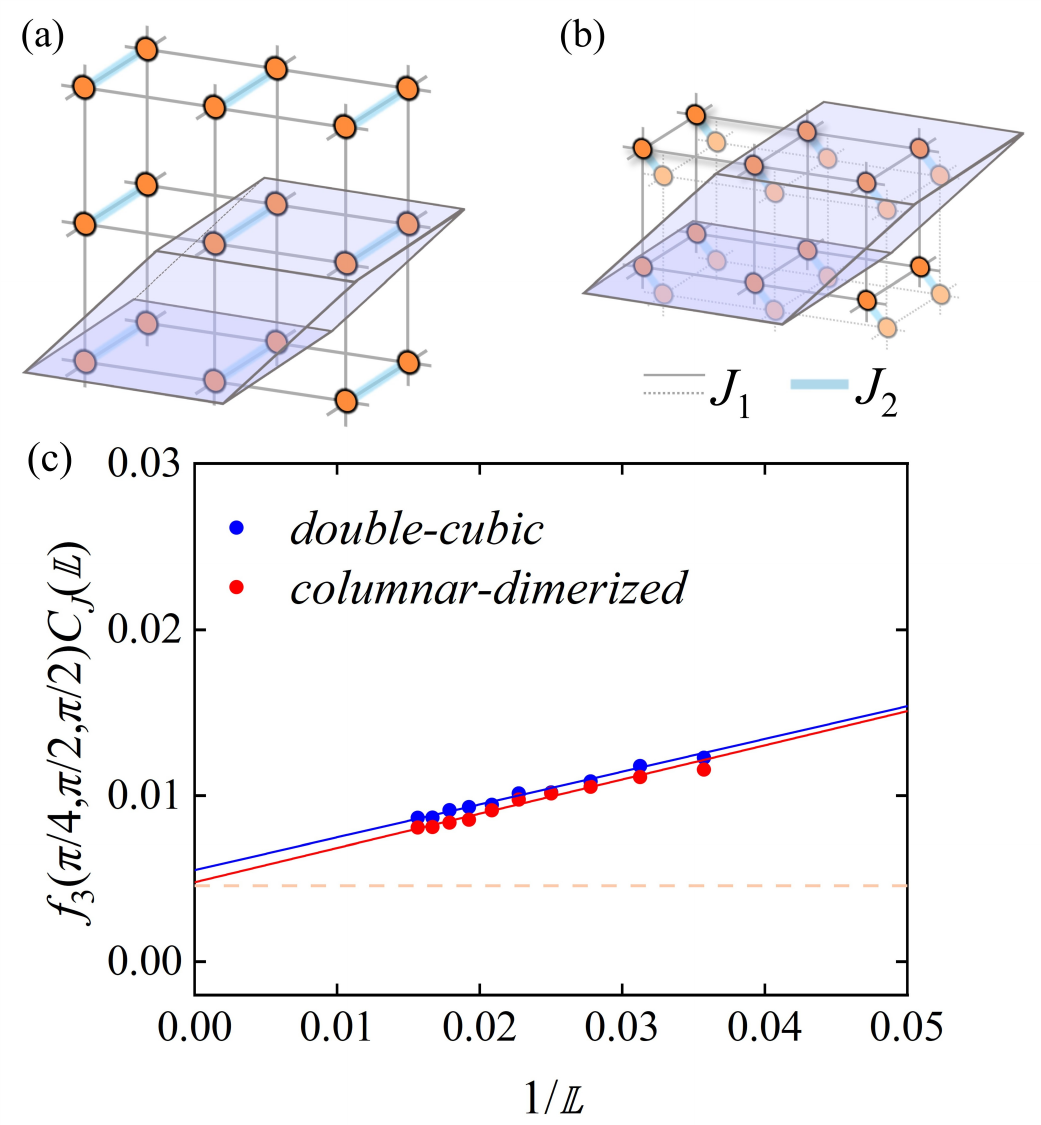}
   \caption{(a) Columnar-dimerized (CD) and (b) double-cubic (DC) Heisenberg models with periodic boundary conditions used in quantum Monte Carlo (QMC) simulations. Thick and thin bonds denote strong and weak exchange couplings. The blue parallelepiped of side length $L$ marks the region where the disorder operator is defined. (c) Finite-size extrapolation of $f_{3}(\pi/4,\pi/2,\pi/2)C_J$ from QMC simulations of the CD and DC models for system sizes up to $\mathds{L}=64$ ($\sim5\times10^5$ spins). The orange dashed line shows the theoretical value.}
    \label{fig:model_CJ}
\end{figure}

 
 \mysec{Monte Carlo for 3+1D O(3) transition.}  We consider the columnar-dimerized (CD) and double-cubic (DC) antiferromagnetic (AFM) Heisenberg models, as shown in FIG.~\ref{fig:model_CJ} (a) and (b). The Hamiltonian of the CD model is
\begin{flalign}
   H_{CD}=J_1\sum\limits_{\left \langle i,j\right \rangle}{{{S}}_{i} \cdot {{S}}_{j}}+J_2\sum\limits_{\left \langle i,j \right \rangle'}{{S}}_{i} \cdot {{S}}_{j}, \nonumber
\end{flalign}
where $\boldsymbol{S}_{i}$ denotes the spin-$1/2$ operator at site $i$. The symbols $\langle i,j\rangle$ and $\langle i,j\rangle'$ label two types of nearest-neighbor bonds with coupling strengths $J_1$ (weak) and $J_2$ (strong), respectively. The quantum critical point has been revealed to $J_2/J_1$=4.0159(1)~\cite{liang2025scaling}.
For the DC model, the Hamiltonian reads
\begin{align}
H_{DC}=J_1\sum\limits_{\left \langle i,j\right \rangle}({{{S}}_{1,i} \cdot {{S}}_{1,j}}+{{{S}}_{2,i} \cdot {{S}}_{2,j}})+J_2\sum\limits_{i}{{S}}_{1,i} \cdot {{S}}_{2,i}, \nonumber
\end{align}
where $J_2$ denotes inter-cube strong AFM Heisenberg coupling. As the coupling ratio $J_2/J_1$ increases, the system undergoes a continuous transition
from the AFM N\'{e}el phase to a spin-singlet phase. The quantum critical point has been determined to be at $J_2/J_1$=4.83704(6)~\cite{qin2015multiplicative}. In practice, we set $J_1= 1$ as the unit of energy and adjust $J_2$ to access the critical point.

In the numerical simulations, we evaluate the disorder operator Eq.~\eqref{eq:_dis_op} through $\mathcal{U}_{\Sigma}(\chi)=\prod_{j\in \Sigma}e^{\mathtt{i}\chi n_{j}}$, where $n_j=S_j^z-\tfrac{1}{2}$ is the $\mathrm{U}(1)$ density operator in the spin-$1/2$ model. The cubic geometry of $\Sigma$ was studied in Ref.~\cite{liang2025scaling}. In this work, we consider a parallelepiped with $\theta_{1,2}=\pi/4$ and $\theta_{1,3}=\theta_{2,3}=\pi/2$ in order to test
the prediction of Eq.~\eqref{eq:_angle_fun_3d-II}. We fit the data using the ansatz
\begin{flalign}
|\langle\mathcal{U}_{\Sigma}(\chi)\rangle|\sim e^{-a(\chi)L^{2}+b(\chi)L+s(\chi)\log L}.
\label{eq:disorder}
\end{flalign}
In the small-$\chi$ limit, the coefficient $s(\chi)$ of the logarithmic term is predicted to take the form 
$s(\chi)=-f_{3}C_J\chi^2/2$. Here, we take the value $C_{J}=1/(4\pi^{4})$ at the Gaussian fixed point~\cite{petkou1996conserved,1994implications,diab2016on}, since the marginally irrelevant interaction at the upper critical dimension does not alter the coefficient of the logarithmic term~\cite{liang2025scaling}.

As shown in the SM~\cite{SM}, the QMC data are quantitatively well described by the scaling form in Eq.~\eqref{eq:disorder}. To access the universal value in the thermodynamic limit, we perform finite-size extrapolations of $f_{3}(\theta_{i,j})C_J$ for both the CD and DC models, as shown in FIG.~\ref{fig:model_CJ} (c). For the CD model, the extrapolated value $f_{3}(\pi/4,\pi/2,\pi/2)C_J=0.0048(5)$ agrees with the theoretical prediction of the mean-field theory $(1+\pi/4)/(4\pi^{4})=0.0046$. For the DC model, the extrapolated value is $0.0055(7)$, showing a modest deviation from the theoretical prediction. This deviation likely originates from stronger finite-size effects and the larger leading quadratic contribution. Overall, the numerical results provide strong support for Eq.~\eqref{eq:_angle_fun_3d-II}.

\begin{figure}
    \centering
 \includegraphics[width=0.8\linewidth]{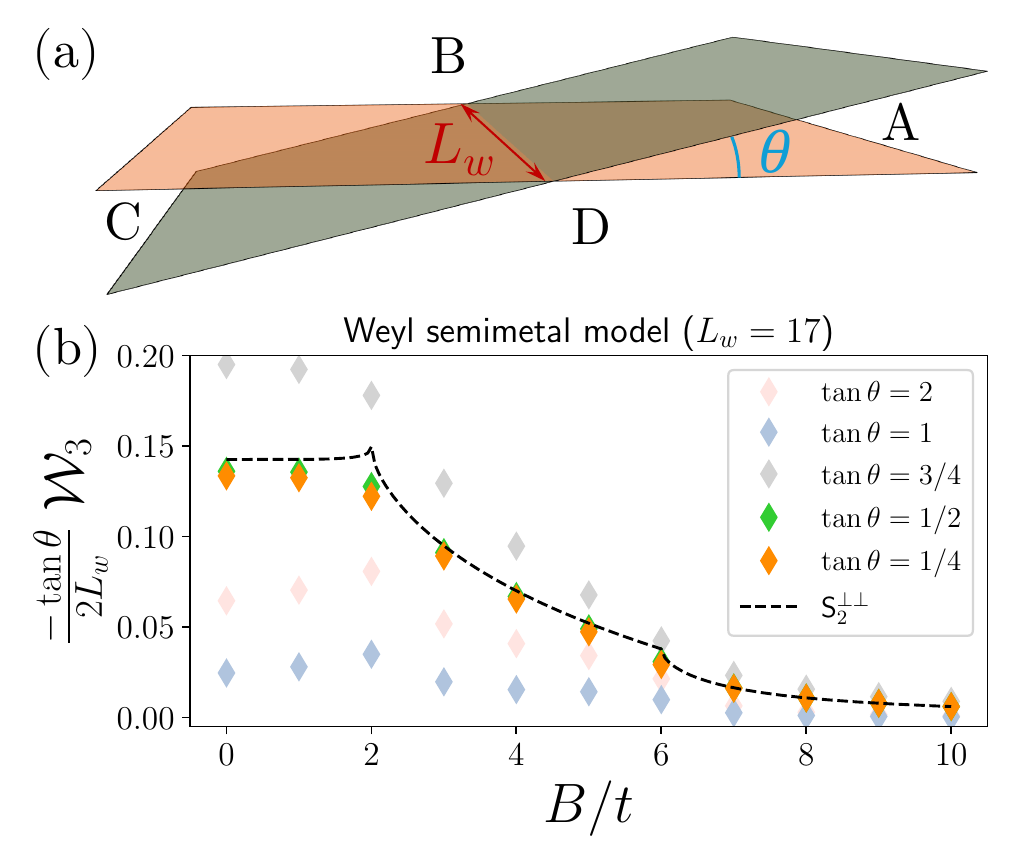}
    \caption{(a) Real-space partition for extracting the coefficient $\mathsf{S}_{2}$. (b) Comparison of $\mathsf{S}_{2}$ obtained from the exact Brillouin-zone integral of the band quantum metric Eq.~\eqref{eq:_QM_Bloch} and that extracted from the wedge-corner contribution, see Eq.~\eqref{eq:_QM_wedge-II}, in a two-band Weyl semimetal model. One of the partition planes are taken to be parallel to a cubic lattice plane.
    }
    \label{fig:_wedge}
\end{figure}

\mysec{Measuring quantum geometry.} Now let us shift gears to a different generalization of the two-dimensional corner charge fluctuations. We consider the real-space partition shown in FIG.~\ref{fig:_wedge}. (a), where two rectangular planes intersect along a one-dimensional wedge that is perpendicular to their edges. We then define the combination
\begin{flalign}
\mathcal{W}_{3}&=\mathcal{F}_{\textrm{A}}+\mathcal{F}_{\textrm{B}}+\mathcal{F}_{\textrm{C}}+\mathcal{F}_{\textrm{D}}-\mathcal{F}_{\textrm{AB}}\nonumber\\&-\mathcal{F}_{\textrm{CD}}-\mathcal{F}_{\textrm{BC}}-\mathcal{F}_{\textrm{AD}}+\mathcal{F}_{\textrm{ABCD}}.
\label{eq:_wedge_scheme_3d}
\end{flalign}
A direct parallel with the two-dimensional analysis~\cite{estienne2022cornering,Tam2024Corner,Wu2025Corner} shows that $\mathcal{W}_{3}=2(\Xi_{\textrm{AC}}+\Xi_{\textrm{BD}})$, where the paring $\Xi_{\textrm{AC}}$ is defined in the same way as in Eq.~\eqref{eq:_SSF_pair_3d}. Since the two regions $\textrm{A}\;(\textrm{B})$ and $\textrm{C}\;(\textrm{D})$ share a common wedge, $\mathcal{W}_{3}$  contains a term proportional to the wedge length
$L_{w}$.

   

As we clarify in the SM~\cite{SM}, for rotationally invariant systems we have the limiting relation 
\begin{flalign}
\lim_{L_{w}\rightarrow\infty}\frac{\mathcal{W}_{3}}{L_{w}}=\frac{2}{\pi}\mathsf{S}_{2}f_{2}(\theta),
\label{eq:_QM_wedge-I}
\end{flalign}
where $f_{2}(\theta)$ is the angle function defined in Eq.~\eqref{eq:_angle_fun_2d}. Here, $\mathsf{S}_{2}$ is the coefficient of the $|\boldsymbol{k}|^{2}$ term in the long-wavelength expansion of the SSF $S(\boldsymbol{k})$\footnote{In gapped insulators, $\mathsf{S}_{2}$ is identified with the many-body quantum metric defined via flux insertion or twisted boundary conditions~\cite{SWM2000,Resta2002Rev,Resta2011Rev,Onishi2024Bound2,Wu2025Corner,Tam2024Corner,Onishi2025quantum} (see also the SM~\cite{SM}). For non-interacting Bloch systems, whether gapped or gapless, $\mathsf{S}_{2}$ is given by the Brillouin-zone integral of the band quantum metric, see Eq.~\eqref{eq:_QM_Bloch}.}. For anisotropic systems, the coefficient becomes a tensor, so that  $S(\boldsymbol{k})\supset\mathsf{S}_{2}^{ij}k_{i}k_{j}$. Generalizing the arguments in Refs.~\cite{Wu2025Corner,Tam2024Corner} for corner charge fluctuations in anisotropic two-dimensional systems \cite{SM}, we find
\begin{flalign}
\mathsf{S}_{2}^{\perp\perp}=\lim_{\theta\rightarrow0}\lim_{L_{w}\rightarrow\infty}\frac{-\theta\mathcal{W}_{3}}{2L_{w}}
\label{eq:_QM_wedge-II}
\end{flalign}
where $\mathsf{S}_{2}^{\perp\perp}$ denotes the component of $\mathsf{S}_{2}^{ij}$ perpendicular to both the wedge and the bipartition plane.

For non-interacting Bloch electrons, $\mathsf{S}_{2}^{ij}$ is simply the integrated quantum metric over the Brillouin zone (BZ)
\begin{flalign}
\mathsf{S}_{2}^{ij}=\int_{\textrm{BZ}}\frac{\textrm{d}^{3}\boldsymbol{k}}{(2\pi)^{3}}g^{ij}(\boldsymbol{k}).
\label{eq:_QM_Bloch}
\end{flalign}
Here, the band quantum metric $g^{ij}$ is given by $g^{ij}(\boldsymbol{k})=\frac{1}{2}\textrm{Tr}(\partial^{i}P\partial^{j}P)$, where $P$ denotes the projector onto the occupied bands and $\partial^i\equiv\partial/\partial k_i$. In three dimensions, Eq.~\eqref{eq:_QM_Bloch} becomes dimensionful and sensitive to the lattice constant, but remains finite in both gapped and gapless systems. As a simple illustration, we consider the two-band Weyl-semimetal Hamiltonian~\cite{Mccormick2017Weyl}, $\mathcal{H}_{\boldsymbol{k}}=\vec{d}_{\boldsymbol{k}}\cdot\vec{\sigma}$, where $\vec{\sigma}$ is the $3$-vector of Pauli matrices and $\vec{d}_{\boldsymbol{k}}=(-2t\sin k_{1},-2t\sin k_{2},-B-2t\sum_{i=1}^{3}\cos k_{i})$. We shall focus on $B/t>0$ and fill the lower band. For $B/t<2$, this model hosts two pairs of Weyl points positioned at $(k_1, k_2)=(0,\pi)$ and $(\pi,0)$, with $k_3=\pm \arccos(-B/2t)$. The presence of Weyl points is associated with an enhanced quantum metric. For $2< B/t<6$, there is one pair of Weyl points split at $\boldsymbol{k}=(\pi, \pi, \pm \arccos(2-B/2t))$. The system is gapped for $B/t>6$ and the quantum metric is suppressed. 
In FIG.~\ref{fig:_wedge}(b), we compare the integrated quantum metric extracted from wedge charge fluctuations $\mathcal{W}_3$ and the exact value determined by $g^{ij}(\boldsymbol{k})=\frac{1}{4}\partial^{i}\hat{d_{\boldsymbol{k}}}\cdot\partial^{j}\hat{d_{\boldsymbol{k}}}$ where $\hat{d_{\boldsymbol{k}}}=\vec{d}_{\boldsymbol{k}}/|\vec{d}_{\boldsymbol{k}}|$. The numerical results with small wedge angles support Eq.~\eqref{eq:_QM_wedge-II}.

   

\mysec{General dimensions.} Given the structural similarity between Eq.~\eqref{eq:_Int_Gram_2d} and Eq.~\eqref{eq:_Int_Gram_3d}, it is natural to expect that Eq.~\eqref{eq:_angle_fun_2d} and Eq.~\eqref{eq:_angle_fun_3d-I} admit a unified generalization that applies to arbitrary dimensions. As shown in the SM~\cite{SM}—where a method complementary to the subtraction schemes in FIG.~\ref{fig:_corner} is used to directly extract the dimensionless contribution $\mathcal{C}_{d}$ to the bipartite fluctuations of a $d$-dimensional parallelotope—the resulting general formula takes the form $\mathcal{C}_{d}=f_{d}(\theta_{i,j})\varUpsilon_{d}$, where the $d$-th radial moment of the SSF is 
\begin{flalign}
\varUpsilon_{d}=-\int_{a}^{L}\textrm{d}rr^{2d-1}S(r)
\label{eq:_dth_mom}
\end{flalign}
and the generalized universal angle function reads
\begin{flalign}
f_{d}(\theta_{i,j})=(-1)^{d+1}\sum_{\{\mathbf{s}=\pm\}}I_{d}(G^{\{\mathbf{s}\}}).
\label{eq:_angle_fun_dd-I}
\end{flalign}
Here $I_d$ generalizes Eq.~\eqref{eq:_Int_Gram_2d} and Eq.~\eqref{eq:_Int_Gram_3d}.
A $d$-dimensional parallelotope is specified by a set of linearly independent unit vectors $\hat{\boldsymbol{e}}_{1},\hat{\boldsymbol{e}}_{2},\ldots,\hat{\boldsymbol{e}}_{d}$, which define the directions of its edges. The Gram matrix appearing in Eq.~\eqref{eq:_angle_fun_dd-I} is defined as in Eq.~\eqref{eq:_Gram_matrix} using these vectors. The dependence on the $d(d-1)/2$ opening angles $\theta_{i,j}$ between $\hat{\boldsymbol{e}}_{i}$ and $\hat{\boldsymbol{e}}_{j}$ (where $i<j$) is encoded in the Gram-matrix functional
\begin{flalign}
I_{d}(G)=\det(G)\int_{\Delta_{d-1}}\textrm{d}^{d-1}v\frac{\prod_{i=1}^{d}v_{i}}{(v^{\mathsf{T}}Gv)^{d}},
\label{eq:_Int_Gram_dd}
\end{flalign}
where $\Delta_{d-1}$ is the $(d-1)$-simplex defined by $\sum_{i=1}^{d}v_{i}=1$ and $v_{i}\geq0$ for $i=1,\ldots,d$. For the special configuration with $\theta_{1,2}=\theta$ and $\theta_{i,j}=\pi/2$ for all other pairs $i<j$, the angle function simplifies to
\begin{flalign}
f_{d}(\theta,\pi/2,\ldots,\pi/2)=\frac{(-1)^{d}}{\Gamma(d)}f_{2}(\theta),
\label{eq:_angle_fun_dd-II}
\end{flalign}
where $f_{2}(\theta)$ is given by Eq.~\eqref{eq:_angle_fun_2d}.

The radial moment $\varUpsilon_{d}$ in Eq.~\eqref{eq:_dth_mom} provides a dimensionless measure of the spatial locality of charge correlations. For gapped insulators, where correlations decay exponentially, Eq.~\eqref{eq:_dth_mom} converges to a finite value. (When $d=2$, it reduces to the many-body quantum metric~\cite{Wu2025Corner,Tam2024Corner}.)
For CFTs in any dimension, the power-law decay $S(r)=-C_{J}/r^{2d}$ leads to a universal logarithmic divergence
$\varUpsilon_{d}=C_{J}\log(L/a)$,
where $C_{J}$ is the current central charge, which is related to the zero-temperature critical conductivity via
$\mathrm{Re}\,\sigma(\omega)\propto C_{J}\omega^{d-2}$~\cite{Lucas2017CFTRes}.

Metallic systems require a more nuanced discussion.
Because charge correlations decay more slowly than in CFTs, the leading boundary-law term in the bipartite fluctuations is enhanced to $L^{d-1}\log L$~\cite{FSEE1,FSEE2,FSEE_swingle1,FSEE_swingle2,FSEE_Ryu,Cai_FS_2024,Wu2025CFS}.
For $d=2$, the subleading contribution $\mathcal{C}_{2}$ scales linearly with $L$, as shown in Ref.~\cite{estienne2022cornering}, and is therefore qualitatively distinct from the universal logarithmic term found in CFTs. In contrast, in $d=3$, a $|\boldsymbol{k}|^{3}$-term generically appears in the SSF $S(\boldsymbol{k})$ with a dimensionless coefficient that is independent of the Fermi momentum $k_{F}$.
This term mirrors the scaling form found in CFTs (see the SM~\cite{SM}) and gives rise to a logarithmic contribution to $\mathcal{C}_{3}$. More generally, explicit analyses of (rotationally invariant) Fermi gases and Landau Fermi liquids reveal a characteristic even-odd dependence on the spatial dimension $d$, as shown in the SM~\cite{SM}.
These results are summarized in Table~\ref{tab:_dth_mom}.

The wedge-based method for extracting the coefficient $\mathsf{S}_{2}$ extends naturally to arbitrary dimensions. We consider the real-space partition in FIG.~\ref{fig:_wedge}, where the one-dimensional interval $[0,L_{w}]$ is replaced by a $(d-2)$-dimensional cube $[0,L_{w}]^{d-2}$. Defining $\mathcal{W}_{d}$ using the same expression as Eq.~\eqref{eq:_wedge_scheme_3d}, the quantity $\mathcal{W}_{d}/L_{w}^{d-2}$ satisfies the same limiting relation as Eq.~\eqref{eq:_QM_wedge-I}. For anisotropic systems, the generalization of Eq.~\eqref{eq:_QM_wedge-II} is 
\begin{flalign}
\mathsf{S}_{2}^{\perp\perp}=\lim_{\theta\rightarrow0}\lim_{L_{w}\rightarrow\infty}\frac{-\theta\mathcal{W}_{d}}{2L_{w}^{d-2}}.
\label{eq:_QM_wedge-III}
\end{flalign}

\begin{table}
\begin{tabular}{|c|c|c|c|}
\hline 
 & Insulators & CFTs & Metals\tabularnewline
\hline 
$d=\textrm{odd}$ & finite & $\log L$ & $\log L+L^{2}+L^{4}+\ldots+L^{d-1}$\tabularnewline
\hline 
$d=\textrm{even}$ & finite & $\log L$ & $L+L^{3}+L^{5}+\ldots+L^{d-1}$\tabularnewline
\hline 
\end{tabular}

\caption{Large-$L$ scaling behavior of the $d$-th radial moment defined in Eq.~\eqref{eq:_dth_mom} for gapped insulators, conformal field theories (CFTs), and Fermi-liquid metals in spatial dimension $d$.}

\label{tab:_dth_mom}

\end{table}

\mysec{Summary and discussion.} In this work, we derive the universal angle function governing bipartite fluctuations in general spatial dimensions $d$, applicable to a broad class of many-body systems. We further establish a general protocol for extracting quantum geometry from charge fluctuations. In three dimensions, our analytical predictions are corroborated by detailed numerical simulations of lattice models.

Although our simulations focus on the conventional $\mathrm{O}(3)$ symmetry-breaking transition, they serve as a proof of principle for the applicability of our framework to CFTs in $d=3$. In particular, our results provide a concrete numerical diagnostic for candidate lattice realizations of unconventional quantum critical points, such as those proposed in Ref.~\cite{Bi_4D_2019}, and thereby motivate future large-scale numerical studies in higher dimensions.

More broadly, this work is inspired by the growing connections between bipartite fluctuations, disorder operators, and entanglement entropy (EE)~\cite{EEBF1,EEBF2,EEBF3,EEBF4,EEBF5,EEBF6,EEBF7,EEBF8,EEBF9,EEvsBF,liu2022multipart,liu2024multipart,JRZhao2022,JRZhao2021,song2024quantum,song2023extracting,deng2023improved,deng2024diagnosing,wang2024ee,wang2024probing,ding2024tracking,jiang2024high}. Recent studies of EE in CFTs have identified logarithmic contributions associated with trihedral corners in $d=3$~\cite{4D_corner_1,4D_corner_2,4D_corner_3,4D_corner_4,4D_corner_5,EE_singular_2019}. Our results raise the question of whether EE in CFTs exhibits the same universal angle dependence identified here, which we leave for future investigation.

Our results suggest several natural extensions. First, in $d=2$ the universal angle function has been generalized to anisotropic systems~\cite{Wu2025Corner}, where the corner contribution depends on both the opening angle and the absolute orientation of the subregion. An analogous generalization in higher dimensions should be possible, for example through spherical-harmonic expansions of the corner term. Second, while the logarithmic term in CFTs is controlled by $C_{J}$, similar logarithmic behavior can also arise at certain metallic critical points without conformal symmetry in $d=2$~\cite{Wu2025CFS}. It would be interesting to explore whether analogous phenomena occur at critical points in $d=3$, such as the example in Ref.~\cite{cmit3d}. Finally, we focus on the second cumulant Eq.~\eqref{eq:_2nd_cumulant} of the disorder operator. Extending the analysis to higher-order cumulants would provide a higher-dimensional generalization of Refs.~\cite{FCS_Corner,Tam2024Corner}. Given the precise relation between cumulants and EE for free fermions~\cite{EEBF1,EEBF2,EEBF3,EEBF4,EEBF5,EEBF6,EEBF7,EEBF8,EEBF9,EEvsBF}, such results may also shed light on corner contributions to EE. It would be particularly interesting to explore the relation between corner EE and quantum geometry in three dimensions or above. Finally, given the advancement of quantum gas microscopy for measuring real-space correlations \cite{nelson2007imaging,3DQGM2020,Bakr_review}, our work shall provide the theoretical basis for probing quantum criticality and quantum geometry in future quantum simulators.

\mysec{Acknowledgments.}
We are thankful to Meng Cheng, Jonah Herzog-Arbeitman and Jiabin Yu for inspiring discussions. We also thank Duncan Haldane and Yarden Sheffer for discussions and collaboration on a related project.
X.W. and S.R. were supported by the Simons Investigator Grant (566116). P.M.T. was supported by a postdoctoral fellowship at the Princeton Center for Theoretical Science and a Croucher Fellowship for Postdoctoral Research.
S.R. acknowledges support from the Gordon and Betty Moore Foundation EPiQS initiative, Grant GBMF8685.01. X.L. and D.X.Y. were supported by NKRDPC-2022YFA1402802, Research Center for Magnetoelectric Physics of Guangdong Province (Grant No. 2024B0303390001). Z.L. acknowledges support from the China Postdoctoral Science Foundation under Grant No.2024M762935.

\mysec{Note added:} We would like to draw the reader’s attention to a related paper~\cite{FS_Geom_2026}, to appear on the same arXiv listing, which investigates the logarithmic contribution to charge fluctuations in three-dimensional metals for a spherical (or ellipsoidal) partition surface and its connection to Fermi-surface geometry, complementary to the parallelepiped geometry considered here.


\let\oldaddcontentsline\addcontentsline
\renewcommand{\addcontentsline}[3]{}
\bibliography{Ref}
\let\addcontentsline\oldaddcontentsline

\clearpage
\newpage
\onecolumngrid

\appendix

\begin{center}
\textbf{\large Supplemental Materials for ``Corner Charge Fluctuations in Higher Dimensions''}\\
\end{center}

\setcounter{secnumdepth}{2}

\tableofcontents

\section{Many-Body Quantum Geometry}
\label{supp_sec:_Quan_Geom}

In this section, we collect a few relevant background results on many-body quantum geometry~\cite{SWM2000,Resta2002Rev,Resta2011Rev,Onishi2024Bound2,Wu2025Corner,Tam2024Corner,Onishi2025quantum} for the reader’s convenience. We study a $d$-dimensional system with periodic boundary conditions, where the system spans lengths $L_{1},\ldots,L_{d}$ along the $d$ spatial directions, forming a $d$-dimensional torus. In the presence of background fluxes $\varPhi_{i}\in[0,2\pi)$, the gauge potential takes the form  $A_{i}=-\varPhi_{i}/L_{i}$ along the $i$-th direction. Using the ground-state wave functions $|\psi_{m}\rangle$, the many-body quantum geometric tensor is defined as 
\begin{flalign}
\mathcal{Q}_{mn}^{ij}(\boldsymbol{A})=\frac{1}{V}\left\langle \!\frac{\partial\psi_{m}}{\partial A_{i}}\!\right|\!(1-\mathcal{P})\!\left|\!\frac{\partial\psi_{n}}{\partial A_{j}}\!\right\rangle,
\end{flalign}
where $V=\prod_{i=1}^{d}L_{i}$ is the system volume, $m,n$ label the (possibly degenerate) ground states, and $\mathcal{P}=\sum_{n=1}^{\textrm{deg}}|\psi_{n}\rangle\langle\psi_{n}|$ denotes the projection onto the many-body ground-state manifold. It can be decomposed as $\mathcal{Q}^{ij}=\mathcal{G}^{ij}-\frac{\mathtt{i}}{2}\mathcal{F}^{ij}$, where $\mathcal{G}^{ij}$ and $\mathcal{F}^{ij}$ represent the (non-Abelian) quantum metric and Berry curvature, respectively. For gapped systems on a torus, both quantities are independent of the background fluxes $\boldsymbol A$ (up to exponentially small finite-size corrections) and are therefore constant on the flux torus~\cite{niu1985quantized,SWM2000,kudo2019many}.

For a generic interacting gapped insulator, the static structure factor obeys the long-wavelength expansion
\begin{flalign}
S(\boldsymbol{q})=S(-\boldsymbol{q})=\mathsf{S}_{2}^{ij}q_{i}q_{j}+\mathsf{S}_{4}^{ijkl}q_{i}q_{j}q_{k}q_{l}+\ldots
\end{flalign}
where each coefficient $\mathsf{S}_{2n}$ is a tensor determined by the microscopic properties of the system. The coefficient $\mathsf{S}_{2}^{ij}$ is fully determined by the quantum geometry, and we can express it as
\begin{flalign}
\mathsf{S}_{2}^{ij}=\sum_{n=1}^{\textrm{deg}}\frac{\mathcal{G}_{nn}^{ij}(\boldsymbol{A})}{\textrm{deg}}=\frac{1}{\textrm{deg}}\frac{1}{V}\frac{1}{2}\textrm{Tr}\left(\frac{\partial\mathcal{P}}{\partial A_{i}}\frac{\partial\mathcal{P}}{\partial A_{j}}\right).
\label{eq:_SSF_and_QM}
\end{flalign}
The relation between $\mathsf{S}_{2}^{ij}$ and $\mathcal{G}^{ij}$ can be established using the Souza-Wilkens-Martin (SWM) sum rule~\cite{SWM2000}
\begin{flalign}
\mathsf{S}_{2}^{ij}=\int_{0}^{+\infty}\frac{\textrm{d}\omega}{\pi}\frac{\textrm{Re}\sigma_{+}^{ij}(\omega)}{\omega},
\end{flalign}
where $\sigma_{+}^{ij}=(\sigma^{ij}+\sigma^{ji})/2$ denotes the longitudinal component of the conductivity tensor. Nontrivial ground-state degeneracy ($\mathrm{deg}>1$) occurs in topologically ordered phases such as fractional quantum Hall states, while a generic non-interacting band insulator at integer filling has a unique many-body ground state ($\mathrm{deg}=1$).

To connect the many-body formulation with single-particle quantum geometry, let us consider a many-body ground state that takes the form of a Slater determinant of Bloch orbitals. In this case, one can show that the many-body quantum metric reduces to the Brillouin-zone integral of the band quantum metric
\begin{equation}
\mathcal{G}^{ij}(\boldsymbol{A})=\frac{1}{V}\sum_{\boldsymbol{k}}g^{ij}(\boldsymbol{k})=\int_{\textrm{BZ}}\frac{\textrm{d}^{d}\boldsymbol{k}}{(2\pi)^{d}}g^{ij}(\boldsymbol{k}),
\end{equation}
where $V$ is the total volume of the system, and $g^{ij}(\boldsymbol{k})$
denotes the band quantum metric
\begin{equation}
g^{ij}(\boldsymbol{k})=\frac{1}{2}\textrm{Tr}\left[\frac{\partial P(\boldsymbol{k})}{\partial k_{i}}\frac{\partial P(\boldsymbol{k})}{\partial k_{j}}\right],
\end{equation}
with $P(\boldsymbol{k})$ the projector onto the occupied bands. This expression remains valid in the presence of degenerate bands. This formulation is fully consistent with the static structure factor of a non-interacting insulator, which is given by
\begin{flalign}
S(\boldsymbol{q}) & =\int_{\textrm{BZ}}\frac{\textrm{d}^{d}\boldsymbol{k}}{(2\pi)^{d}}\textrm{Tr}[P(\boldsymbol{k})\bar{P}(\boldsymbol{k}+\boldsymbol{q})]=\int_{\textrm{BZ}}\frac{\textrm{d}^{d}\boldsymbol{k}}{(2\pi)^{d}}\textrm{Tr}[P(\boldsymbol{k})(P(\boldsymbol{k})-P(\boldsymbol{k}+\boldsymbol{q}))],
\end{flalign}
where $\bar{P}(\boldsymbol{k})=1-P(\boldsymbol{k})$ is the complement
projector. It
admits the long-wavelength expansion
\begin{equation}
S(\boldsymbol{q})=-\int_{\textrm{BZ}}\frac{\textrm{d}^{d}\boldsymbol{k}}{(2\pi)^{d}}\textrm{Tr}\left[P(\boldsymbol{k})\sum_{n=1}^{+\infty}\frac{(\boldsymbol{q}\cdot\partial_{\boldsymbol{k}})^{2n}}{(2n)!}P(\boldsymbol{k})\right]=\mathsf{S}_{2}^{ij}q_{i}q_{j}+\mathsf{S}_{4}^{ijkl}q_{i}q_{j}q_{k}q_{l}+\ldots.
\end{equation}
The quadratic term is determined by the integrated quantum metric
\begin{equation}
\mathsf{S}_{2}^{ij}=\int_{\textrm{BZ}}\frac{\textrm{d}^{d}\boldsymbol{k}}{(2\pi)^{d}}\frac{1}{2}\textrm{Tr}\left[\frac{\partial P(\boldsymbol{k})}{\partial k_{i}}\frac{\partial P(\boldsymbol{k})}{\partial k_{j}}\right]=\int_{\textrm{BZ}}\frac{\textrm{d}^{d}\boldsymbol{k}}{(2\pi)^{d}}g^{ij}(\boldsymbol{k})=\mathcal{G}^{ij}(\boldsymbol{A}),
\end{equation}
in agreement with Eq. \eqref{eq:_SSF_and_QM}. More generally,
the rank-$2n$ symmetric tensor is given by
\begin{equation}
\mathsf{S}_{2n}^{i_{1}\ldots i_{n}j_{1}\ldots j_{n}}=\frac{(-1)^{n-1}}{(2n)!}\int_{\textrm{BZ}}\frac{\textrm{d}^{d}\boldsymbol{k}}{(2\pi)^{d}}\textrm{Tr}\left[\frac{\partial^{n}P(\boldsymbol{k})}{\partial k_{i_{1}}\ldots\partial k_{i_{n}}}\frac{\partial^{n}P(\boldsymbol{k})}{\partial k_{j_{1}}\ldots\partial k_{j_{n}}}\right].
\end{equation}

\section{The $d$-th Radial Moment of SSF}

In this section, we analyze the $d$-th radial moment $\varUpsilon_{d}$ defined in Eq.~\eqref{eq:_dth_mom} for the static structure factor (SSF) $S(r)$ of rotationally invariant systems. We begin with several general observations without specifying the explicit form of $S(r)$. The Fourier transform of the static structure factor reads
\begin{flalign}
S(\boldsymbol{k})=\int\textrm{d}^{d}\boldsymbol{r}S(\boldsymbol{r})e^{-\mathtt{i}\boldsymbol{k}\cdot\boldsymbol{r}}&=\int_{0}^{+\infty}\textrm{d}r\int_{0}^{\pi}\textrm{d}\theta\Omega_{d-2}(\sin\theta)^{d-2}e^{-\mathtt{i}|\boldsymbol{k}|r\cos\theta}S(r)\nonumber\\&=\int_{0}^{+\infty}\textrm{d}r2\pi^{\frac{d}{2}}r^{d-1}{}_{0}\tilde{\mathrm{F}}_{1}\left(\frac{d}{2},-\frac{r^{2}|\boldsymbol{k}|^{2}}{4}\right)S(r),
\end{flalign}
where $_{0}\tilde{\mathrm{F}}_{1}(a,z)$ is the regularized confluent hypergeometric function. A useful mathematical relation is
\begin{flalign}
\lim_{k\rightarrow0}\left(\frac{\textrm{d}}{\textrm{d}k}\right)^{d}{}_{0}\tilde{\mathrm{F}}_{1}\left(\frac{d}{2},-\frac{r^{2}k^{2}}{4}\right)=\begin{cases}
\displaystyle
\frac{(-1)^{\frac{d}{2}}r^{d}}{2^{d-1}\Gamma(\frac{d}{2})} & d=\textrm{even}\\\\
\displaystyle
0 & d=\textrm{odd}
\end{cases}.
\end{flalign}
When $d$ is even, assuming that $S(k)$ is an analytic and well-behaved function (as in gapped insulators), the $d$-th radial moment Eq.~\eqref{eq:_dth_mom} is directly related to derivatives of $S(k)$  
\begin{flalign}
\int_{0}^{+\infty}\textrm{d}rr^{2d-1}S(r)=\lim_{k\rightarrow0}\frac{2^{d-2}\Gamma(\frac{d}{2})}{(-\pi)^{\frac{d}{2}}}\left(\frac{\textrm{d}}{\textrm{d}k}\right)^{d}S(k).
\end{flalign}
When $d$ is odd, this relation no longer holds automatically. Nevertheless, $\varUpsilon_{d}$ can still be tied to the small-$k$ expansion of $S(k)$, provided that $S(k)$ develops a certain non-analyticity (as in the case of metals).

In the remainder of this section, we examine several explicit examples, covering both gapped and gapless systems.

\subsection{Conformal Field Theories}

In $(d+1)$-dimensional Euclidean flat spacetime, any conformal field theory (CFT) possesses the following two-point correlation function for a conserved current $J_{\mu}$ associated with a (non-)abelian continuous symmetry group $G$~\cite{Poland2019RMP}
\begin{flalign}
&\langle J_{\mu}^{\mathsf{A}}(x)J_{\nu}^{\mathsf{B}}(0)\rangle=\frac{C_{J}}{\left|x\right|^{2d}}\left(\delta^{\mu\nu}-\frac{2x^{\mu}x^{\nu}}{\left|x\right|^{2}}\right)\delta^{\mathsf{AB}},\nonumber\\&\langle J_{\mu}^{\mathsf{A}}(k)J_{\nu}^{\mathsf{B}}(-k)\rangle=-C_{J}\frac{\pi^{\frac{d+1}{2}}\Gamma(\frac{3-d}{2})}{2^{d-2}\Gamma(d+1)}\left|k\right|^{d-1}\left(\delta_{\mu\nu}-\frac{k_{\mu}k_{\nu}}{\left|k\right|^{2}}\right)\delta^{\mathsf{AB}},
\end{flalign}
where $C_{J}$ is the current central charge, and $\mathsf{A}=1,2,\ldots,\textrm{dim}(G)$ labels the generators of  of $G$. In the following, we focus on a U(1) subgroup of $G$. According to the Kubo formula, the universal conductivity is given by~\cite{Lucas2017CFTRes} 
\begin{flalign}
\textrm{Re}\sigma_{xx}(\omega)=C_{J}\frac{2^{2-d}\pi{}^{\frac{d+3}{2}}}{\Gamma(d+1)\Gamma(\frac{d-1}{2})}\omega^{d-2}\quad\textrm{where}\quad d\geq2.
\end{flalign}

The equal-time density-density correlation obeys the power law
\begin{flalign}
S(\boldsymbol{r})=-\langle J_{\tau}(\tau\rightarrow0,\boldsymbol{r})J_{\tau}(0,\boldsymbol{0})\rangle=-\frac{C_{J}}{|\boldsymbol{r}|^{2d}}.
\label{eq:_CFT_SSF-I}
\end{flalign}
Fourier transforming Eq.~\eqref{eq:_CFT_SSF-I} yields the momentum-space scaling 
\begin{flalign}
S(\boldsymbol{k})&=\begin{cases}
\displaystyle
(-1)^{\frac{d+3}{2}}C_{J}\frac{2^{-d}\pi^{1+\frac{d}{2}}}{\Gamma(d)\Gamma(1+\frac{d}{2})}|\boldsymbol{k}|^{d} & d=\textrm{odd}\\\\
\displaystyle
(-1)^{\frac{d+2}{2}}C_{J}\frac{2^{1-d}\pi^{\frac{d}{2}}}{\Gamma(d)\Gamma(1+\frac{d}{2})}|\boldsymbol{k}|^{d}\log(1/|\boldsymbol{k}|) & d=\textrm{even}
\end{cases}\nonumber\\&=\begin{cases}
+\pi C_{J}|\boldsymbol{k}| & d=1\\
+\frac{\pi}{2}C_{J}|\boldsymbol{k}|^{2}\log(1/|\boldsymbol{k}|) & d=2\\
-\frac{\pi^{2}}{12}C_{J}|\boldsymbol{k}|^{3} & d=3\\
-\frac{\pi^{2}}{69}C_{J}|\boldsymbol{k}|^{4}\log(1/|\boldsymbol{k}|) & d=4\\
+\frac{\pi^{3}}{1440}C_{J}|\boldsymbol{k}|^{5} & d=5\\
+\frac{\pi^{3}}{23040}C_{J}|\boldsymbol{k}|^{6}\log(1/|\boldsymbol{k}|) & d=6\\
\qquad\vdots & \vdots
\end{cases}.
\label{eq:_CFT_SSF-II}
\end{flalign}
Due to Eq.~\eqref{eq:_CFT_SSF-I}, the $d$-th moment $\varUpsilon_{d}$ exhibits a logarithmic dependence on the UV cutoff $a$ and the IR cutoff $L$
\begin{flalign}
\varUpsilon_{d}=-\int_{a}^{L}\textrm{d}rr^{2d-1}S(r)=C_{J}\log(L/a).
\end{flalign}
This underlies the universal logarithmic corner terms discussed in the main text.

For any gapped phase that is adiabatically connected to a conformal quantum critical point with current central charge $C_{J}$, the static structure factor exhibits the asymptotic behaviors
\begin{flalign}
S(r)\approx\begin{cases}
-C_{J}r^{-2d} & r/\xi\ll1\\
-C_{0}\xi^{\alpha-2d}r^{-\alpha}\exp(-r/\xi) & r/\xi\gg1
\end{cases},
\label{eq:_Gap_SSF}
\end{flalign}
where $\xi$ is the correlation length and $C_{0}$ is a dimensionless constant. The $d$-th moment can be estimated as
\begin{flalign}
\varUpsilon_{d}=-\int\textrm{d}rr^{2d-1}S(r)\approx\int_{a}^{\xi}\textrm{d}r\frac{C_{J}}{r}+\int_{\xi}^{+\infty}\textrm{d}rC_{0}\frac{r^{2d-1-\alpha}}{\xi^{2d-\alpha}}e^{-r/\xi}=C_{J}\log(\xi/a)+C_{0}\Gamma(2d-\alpha,1),
\label{eq:_Gap_mom}
\end{flalign}
where $a$ denotes a short-distance cutoff, and $\Gamma(s,x)=\int_{x}^{\infty}\textrm{d}tt^{s-1}e^{-t}$ is the incomplete Gamma function. Importantly, the $d$-th moment $\varUpsilon_{d}$ is free of IR divergence. The exponent $\alpha$ generally depends on the specific gapped phase, and explicit examples will be provided below.


\subsection{Massive $\mathrm{O}(N)$ Bosons}

We then consider the free massive $\textrm{O}(N)$ vector model
\begin{flalign}
\mathcal{L}=\frac{1}{2}|(\partial_{\mu}-\mathtt{i}\vec{A}_{\mu}\cdot\vec{T})\boldsymbol{\phi}|^{2}+\frac{m^{2}}{2}|\boldsymbol{\phi}|^{2},
\end{flalign}
where $\vec{A}_{\mu}$ is a background gauge field, and $\vec{T}$
denotes the generators of $\mathrm{O}(N)$ in the fundamental representation.
A convenient basis for the generators $T^{ab}$ (with $1\leq a<b\leq N$) is given by
\begin{equation}
(T^{ab})_{ij}=-\mathtt{i}(\delta_{i}^{a}\delta_{j}^{b}-\delta_{i}^{b}\delta_{j}^{a}),\label{eq:_O(N)_generators}
\end{equation}
which satisfies $(T^{ab})^{\mathsf{T}}=-T^{ba}$, $(T^{ab})^{\dagger}=T^{ab}$, and $\textrm{Tr}(T^{ab}T^{cd})=2\delta^{ac}\delta^{bd}$.
By definition, the conserved current follows from variation with respect to $\vec{A}_{\mu}$
\begin{equation}
\vec{J}_{\mu}=\lim_{A\rightarrow0}\frac{\delta\mathcal{L}}{\delta\vec{A}_{\mu}}=\frac{\mathtt{i}\boldsymbol{\phi}^{\mathsf{T}}\vec{T}(\partial_{\mu}\boldsymbol{\phi})-\mathtt{i}(\partial_{\mu}\boldsymbol{\phi}^{\mathsf{T}})\vec{T}\boldsymbol{\phi}}{2}=\mathtt{i}\boldsymbol{\phi}^{\mathsf{T}}\vec{T}(\partial_{\mu}\boldsymbol{\phi}),
\end{equation}
which in components yields
 $J_{\mu}^{ab}=\phi^{a}\partial_{\mu}\phi^{b}-\phi^{b}\partial_{\mu}\phi^{a}$. It is convenient to relabel the pair $(a,b)$ by a single index $\mathsf{A}=1,2,\ldots,N(N-1)/2$, in which case $\mathrm{Tr}(T^{\mathsf{A}}T^{\mathsf{B}})=2\delta^{\mathsf{AB}}$. The $\mathrm{U}(1)$ subgroup generated by rotations in the $(1,2)$ plane has the conserved current
 $J_{\mu}=\phi^{1}\partial_{\mu}\phi^{2}-\phi^{2}\partial_{\mu}\phi^{1}$.

 The real-space boson propagator is given by $\langle\phi^{a}(x)\phi^{b}(0)\rangle=\delta^{ab}G_{\phi}(x)$
and 
\begin{equation}
G_{\phi}(x)=\int\frac{\textrm{d}^{d+1}p}{(2\pi)^{d+1}}\frac{1}{|p|^{2}+m^{2}}e^{\mathtt{i}p\cdot x}=\frac{\mathrm{K}_{\frac{d-1}{2}}(|m||x|)}{(2\pi)^{\frac{d+1}{2}}}\left(\frac{|m|}{|x|}\right)^{\frac{d-1}{2}},
\label{eq:_boson_propagator}
\end{equation}
where $\mathrm{K}_{\nu}(z)$ is the modified Bessel function of the second kind. The current two-point function is
\begin{flalign}
\langle J_{\mu}^{\mathsf{A}}(x)J_{\nu}^{\mathsf{B}}(0)\rangle&=\textrm{Tr}(T^{\mathsf{A}}T^{\mathsf{B}})(\partial_{\mu}G_{\phi}(x)\partial_{\nu}G_{\phi}(x)-G_{\phi}(x)\partial_{\mu}\partial_{\nu}G_{\phi}(x))\nonumber\\&=-2\delta^{\mathsf{AB}}\left(\frac{G_{\phi}\partial_{r}G_{\phi}}{r}\delta^{\mu\nu}+\frac{x^{\mu}x^{\nu}}{r^{2}}\left(G_{\phi}\partial_{r}^{2}G_{\phi}-\frac{G_{\phi}\partial_{r}G_{\phi}}{r}-(\partial_{r}G_{\phi})^{2}\right)\right),
\end{flalign}
where $r=|x|$. For the $\mathrm{U}(1)$ subgroup, the equal-time static structure factor reduces to 
\begin{flalign}
S(r)=\frac{G_{\phi}\partial_{r}G_{\phi}}{r}=-\frac{2}{(2\pi)^{d+1}}\left(\frac{|m|}{r}\right)^{d}\textrm{K}_{\frac{d-1}{2}}(|m|r)\textrm{K}_{\frac{d+1}{2}}(|m|r).
\end{flalign}
Using the asymptotic expansions of the modified Bessel functions, we obtain
\begin{flalign}
S(r)\approx\begin{cases}
\displaystyle
-\frac{2}{d-1}\frac{1}{\Omega_{d}^{2}}\frac{1}{r^{2d}} & |m|r\ll1\\\\
\displaystyle
-\frac{\exp(-2|m|r)}{4\pi r^{2}}\left(\frac{|m|}{2\pi r}\right)^{d-1} & |m|r\gg1
\end{cases},
\end{flalign}
where $\Omega_{d}=\Gamma(\frac{d+1}{2})/2\pi^{\frac{d+1}{2}}$ denotes the surface area of the unit $d$-sphere. Comparing with the general asymptotic form in Eq.~\eqref{eq:_Gap_SSF}, we identify the mean-field value of the current central charge
\begin{align}
C_{J}=\frac{2}{d-1}\frac{1}{\Omega_{d}^{2}}=\frac{2}{d-1}\left(\frac{\Gamma(\frac{d+1}{2})}{2\pi^{\frac{d+1}{2}}}\right)^{2}
\end{align}
for the $\mathrm{O}(N)$ transition in $d+1$ dimensions, consistent with Refs.~\cite{1994implications,diab2016on}.

\subsection{Massive Dirac Fermions}

Let us also analyze the asymptotic behaviors in Eq.~\eqref{eq:_Gap_SSF} for massive Dirac fermions. The Lagrangian of the free theory with $N_{f}$ flavors is
\begin{flalign}
\mathcal{L}=\sum_{I=1}^{N_{f}}\bar{\psi}_{I}\gamma^{\mu}(\partial_{\mu}-\mathtt{i}A_{\mu}+m)\psi_{I},
\end{flalign}
where the Dirac matrices $\gamma^{\mu}$ satisfy the Clifford algebra
$\gamma^{\mu}\gamma^{\nu}+\gamma^{\nu}\gamma^{\mu}=2\delta^{\mu\nu}\mathbbm{1}$, and the background field $A_{\mu}$ is introduced to probe the global $\mathrm{U}(1)$ symmetry.
The corresponding conserved current is $J_{\mu}=\delta\mathcal{L}/\delta A_{\mu}=-\mathtt{i}\bar{\psi}_{I}\gamma^{\mu}\psi_{I}$. 

The real-space fermion propagator can be obtained using Eq.~\eqref{eq:_boson_propagator}
\begin{align}
    G_{\psi}(x)&=\int\frac{\textrm{d}^{d+1}p}{(2\pi)^{d+1}}\frac{-\mathtt{i}\slashed{p}+m}{|p|^{2}+m^{2}}e^{\mathtt{i}p\cdot x}=(m-\slashed{\partial})G_{\phi}(x)\nonumber\\&=\left(\frac{|m|}{2\pi}\right)^{\frac{d+1}{2}}\frac{1}{|x|^{\frac{d-1}{2}}}\left(\textrm{sgn}(m)\mathrm{K}_{\frac{d-1}{2}}(|m||x|)+\frac{\slashed{x}}{|x|}\mathrm{K}_{\frac{d+1}{2}}(|m||x|)\right),
\end{align}
where $\mathrm{K}_{\nu}(z)$ is the modified Bessel function of the second kind. The current correlation function then follows as
\begin{align}
   \langle J_{\mu}(x)J_{\nu}(0)\rangle&=-\langle\bar{\psi}_{I}(x)\gamma^{\mu}\psi_{I}(x)\bar{\psi}_{J}(0)\gamma^{\nu}\psi_{J}(0)\rangle=N_{f}\textrm{Tr}(\gamma^{\mu}G_{\psi}(x)\gamma^{\nu}G_{\psi}(-x))\nonumber\\&=\left(\frac{|m|}{2\pi}\right)^{\frac{d+1}{2}}\frac{N_{f}\textrm{Tr}(\mathbbm{1})}{|x|^{d-1}}\left(\begin{array}{c}
(\mathrm{K}_{\frac{d-1}{2}}(|m||x|)^{2}+\mathrm{K}_{\frac{d+1}{2}}(|m||x|)^{2})\delta^{\mu\nu}-2\mathrm{K}_{\frac{d+1}{2}}(|m||x|)^{2}x^{\mu}x^{\nu}/|x|^{2}\\
-2\delta_{d,2}\textrm{sgn}(m)\mathrm{K}_{\frac{d-1}{2}}(|m||x|)\mathrm{K}_{\frac{d+1}{2}}(|m||x|)\mathtt{i}\varepsilon^{\mu\nu\sigma}x^{\sigma}/|x|
\end{array}\right),
\end{align}
where we have used the trace identities for Dirac matrices $\textrm{Tr}(\gamma^{\mu}\gamma^{\nu})=\delta^{\mu\nu}\textrm{Tr}(\mathbbm{1})$,
$\textrm{Tr}(\gamma^{\mu}\gamma^{\nu}\gamma^{\sigma})=\mathtt{i}\textrm{Tr}(\mathbbm{1})\varepsilon^{\mu\nu\sigma}\delta_{d,2}$,
and $\textrm{Tr}(\gamma^{\mu}\gamma^{\rho}\gamma^{\nu}\gamma^{\sigma})=(\delta^{\mu\rho}\delta^{\nu\sigma}-\delta^{\mu\nu}\delta^{\rho\sigma}+\delta^{\mu\sigma}\delta^{\nu\rho})\textrm{Tr}(\mathbbm{1})$. The Chern-Simons term $\propto\varepsilon^{\mu\nu\sigma}x^{\sigma}/|x|$ appears only in $d=2$.

The static structure factor is obtained by taking the equal-time limit of $\langle J_{\tau}J_{\tau}\rangle$
\begin{align}
    S(r)=-N_{f}\textrm{Tr}(\mathbbm{1})\left(\frac{|m|}{2\pi}\right)^{d+1}\frac{\mathrm{K}_{\frac{d-1}{2}}(|m|r)^{2}+\mathrm{K}_{\frac{d+1}{2}}(|m|r)^{2}}{r^{d-1}}.
\end{align}
Using the asymptotic expansions of the modified Bessel functions, we obtain
\begin{align}
S(r)\approx\begin{cases}
\displaystyle
-N_{f}\textrm{Tr}(\mathbbm{1})\frac{1}{\Omega_{d}^{2}}\frac{1}{r^{2d}} & |m|r\ll1\\\\
\displaystyle
-N_{f}\textrm{Tr}(\mathbbm{1})\frac{\exp(-2|m|r)}{2}\left(\frac{|m|}{2\pi r}\right)^{d} & |m|r\gg1
\end{cases},
\end{align}
from which we identify the current central charge
\begin{align}
C_{J}=\frac{N_{f}\textrm{Tr}(\mathbbm{1})}{\Omega_{d}^{2}}=N_{f}\textrm{Tr}(\mathbbm{1})\left(\frac{\Gamma(\frac{d+1}{2})}{2\pi^{\frac{d+1}{2}}}\right)^{2},
\end{align}
in agreement with Ref.~\cite{giombi2016on} for Dirac/Weyl points. The radial moment $\varUpsilon_{d}$ is again given by Eq.~\eqref{eq:_Gap_mom}

\subsection{Free Fermi Gases}

As the simplest Fermi-surface state, we consider the spinless free Fermi gas in $d+1$ dimensions, described by
\begin{align}
    \mathcal{L}=\psi^{\dagger}\left(\partial_{\tau}-\mathtt{i}A_{\tau}-\frac{(\nabla-\mathtt{i}\boldsymbol{A})^{2}}{2m}-\mu\right)\psi
\end{align}
where $m$ is the fermion mass, $\mu$ is the chemical potential, and $A_{\mu}$ is the $\mathrm{U}(1)$ background field. For the isotropic dispersion $\epsilon_{\boldsymbol{k}}=\frac{\boldsymbol{k}^{2}}{2m}-\mu$, the equal-time fermion propagator is
\begin{align}
G_{\psi}(\tau\rightarrow0,\boldsymbol{r})=\int\frac{\textrm{d}^{d}\boldsymbol{k}}{(2\pi)^{d}}\Theta(-\epsilon_{\boldsymbol{k}})e^{-\mathtt{i}\boldsymbol{k}\cdot\boldsymbol{r}}=\left(\frac{k_{F}}{2\pi}\right)^{\frac{d}{2}}\frac{\textrm{J}_{\frac{d}{2}}(k_{F}|\boldsymbol{r}|)}{|\boldsymbol{r}|^{\frac{d}{2}}}
\end{align}
where $k_{F}=\sqrt{2m\mu}$ is the Fermi momentum, and $\textrm{J}_{n}(z)$ denotes the Bessel function of the first kind.

The equal-time density-density correlator takes the form 
\begin{align}
    S(r)=-\left(\frac{k_{F}}{2\pi r}\right)^{d}\textrm{J}_{\frac{d}{2}}(k_{F}r)^{2}.
    \label{eq:_FG_SSF-I}
\end{align}
The $d$-th radial moment defined in Eq.~\eqref{eq:_dth_mom} can be evaluated explicitly
\begin{align}
\varUpsilon_{d}=-\int_{0}^{L}\textrm{d}rr^{2d-1}S(r)=\frac{\Gamma(d)^{2}(k_{F}L)^{2d}}{(4\pi)^{d}\Gamma(\frac{d}{2})}{}_{2}\tilde{\textrm{F}}_{3}\left(\frac{d+1}{2},d;\frac{d+2}{2},d+1,d+1;-k_{F}^{2}L^{2}\right),
\label{eq:_FG_mom-I}
\end{align}
where $_{p}\tilde{\textrm{F}}_{q}(a_{1},\ldots a_{p};b_{1},\ldots,b_{q};z)$ denotes the regularized generalized hypergeometric function. For $d=1$, this reduces to $\varUpsilon_{1}=\log(L)/(2\pi^{2})+\mathscr{O}(1)$. For $d>1$, $\varUpsilon_{d}$ exhibits power-law IR divergences in the cutoff $L$, together with a clear even-odd pattern: a logarithmic term appears only when $d$ is odd, 
\begin{align}
\displaystyle
\varUpsilon_{d}=\begin{cases}
\frac{\log(L)}{2\pi^{d}\Gamma(1-\frac{d}{2})^{2}}+\mathscr{O}(1)+\mathscr{O}(L^{2})+\mathscr{O}(L^{4})+\ldots+\mathscr{O}(L^{d-1}) & d=\textrm{odd}\\\\
\displaystyle
\mathscr{O}(L)+\mathscr{O}(L^{3})+\mathscr{O}(L^{5})+\ldots+\mathscr{O}(L^{d-1}) & d=\textrm{even}
\end{cases}.
\label{eq:_FG_mom-II}
\end{align}
The logarithmic term is reminiscent of the CFT scaling.
Using the Fourier relation in Eq.~\eqref{eq:_CFT_SSF-II}, it maps onto the appearance of a $|\boldsymbol{k}|^{d}$ term in the momentum-space expression $S(\boldsymbol{k})$, together with its dimensionless coefficient:
\begin{align}
S(\boldsymbol{k})\supset\frac{(-1)^{d+1}2^{-d}\pi^{-\frac{d}{2}}}{\Gamma(1-\frac{d}{2})\Gamma(d+1)}|\boldsymbol{k}|^{d}=\begin{cases}
+\frac{|\boldsymbol{k}|}{2\pi} & d=1\\
-\frac{|\boldsymbol{k}|^{3}}{96\pi^{2}} & d=3\\
+\frac{|\boldsymbol{k}|^{5}}{5120\pi^{3}} & d=5\\
-\frac{|\boldsymbol{k}|^{7}}{344064\pi^{4}} & d=7\\
+\frac{|\boldsymbol{k}|^{9}}{28311552\pi^{5}} & d=9\\
-\frac{|\boldsymbol{k}|^{11}}{2768240640\pi^{6}} & d=11\\
\qquad\vdots & \vdots
\end{cases}.
\label{eq:_FG_SSF_dimless}
\end{align}

The prediction in Eq.~\eqref{eq:_FG_SSF_dimless} can be directly verified by the exact expression of $S(\boldsymbol{k})$ for $|\boldsymbol{k}|<2k_{F}$, obtained from the Fourier transform of Eq.~\eqref{eq:_FG_SSF-I}
\begin{align}
\frac{S(\boldsymbol{k})}{\langle\rho\rangle}=\frac{d\Gamma(\frac{d}{2})\tilde{k}}{\sqrt{\pi}\Gamma(\frac{d+1}{2})}{}_{2}\textrm{F}_{1}\left(\frac{1}{2},\frac{1-d}{2};\frac{3}{2};\mathit{\tilde{k}}^{2}\right)&=\begin{cases}
\tilde{\mathit{k}} & d=1\\
\frac{2}{\pi}(\sqrt{1-\mathit{\tilde{\mathit{k}}}^{2}}\mathit{\tilde{\mathit{k}}}+\arcsin(\mathit{\tilde{\mathit{k}}})) & d=2\\
\frac{1}{2}\mathit{\tilde{k}}(3-\mathit{\tilde{k}}^{2}) & d=3\\
\frac{1}{3\pi}(2\mathit{\mathit{\tilde{k}}}\sqrt{1-\mathit{\mathit{\tilde{k}}}^{2}}(5-2\mathit{\mathit{\tilde{k}}}^{2})+6\arcsin(\mathit{\tilde{\mathit{k}}})) & d=4\\
\frac{1}{8}\mathit{\mathit{\tilde{k}}}(15-10\mathit{\mathit{\tilde{k}}}^{2}+3\mathit{\mathit{\tilde{k}}}^{4}) & d=5\\
\frac{1}{15\pi}(2\mathit{\mathit{\mathit{\tilde{k}}}}\sqrt{1-\mathit{\mathit{\mathit{\tilde{k}}}}^{2}}(8\mathit{\mathit{\mathit{\tilde{k}}}}^{4}-26\mathit{\mathit{\mathit{\tilde{k}}}}^{2}+33)+30\arcsin(\mathit{\tilde{\mathit{k}}})) & d=6\\
\qquad\qquad\qquad\vdots & \vdots
\end{cases}\nonumber\\&=\begin{cases}
\tilde{k} & d=1\\
\frac{4}{\pi}\tilde{k}-\frac{2}{3\pi}\tilde{k}^{3}+\mathscr{O}(\mathit{\mathit{\tilde{k}}}^{5}) & d=2\\
\frac{3}{2}\tilde{k}-\frac{1}{2}\tilde{k}^{3} & d=3\\
\frac{16}{3\pi}\mathit{\tilde{k}}-\frac{8}{3\pi}\tilde{k}^{3}+\frac{2}{5\pi}\tilde{k}^{5}+\mathscr{O}(\mathit{\mathit{\tilde{k}}}^{7}) & d=4\\
\frac{15}{8}\mathit{\tilde{k}}-\frac{5}{4}\tilde{k}^{3}+\frac{3}{8}\tilde{k}^{5} & d=5\\
\frac{32}{5\pi}\tilde{k}-\frac{16}{3\pi}\tilde{k}^{3}+\frac{12}{5\pi}\tilde{k}^{5}-\frac{2}{7\pi}\tilde{k}^{7}+\mathscr{O}(\mathit{\mathit{\tilde{k}}}^{9}) & d=6\\
\qquad\qquad\qquad\vdots & \vdots
\end{cases},
\label{eq:_FG_SSF-II}
\end{align}
where $\langle\rho\rangle=\frac{\Omega_{d-1}}{(2\pi)^{d}}\frac{k_{F}^{d}}{d}$ is the charge density, and $_{2}\textrm{F}_{1}(a,b;c;z)$ denotes the hypergeometric function. For convenience, we introduced the dimensionless momentum $\tilde{k}=\frac{|\boldsymbol{k}|}{2k_{F}}$.

\subsection{Landau Fermi Liquids}

The long-wavelength behavior of the static structure factor in a Fermi liquid is qualitatively identical to that of free fermions: only odd powers of $|\boldsymbol{k}|$ appear in $S(\boldsymbol{k})$, with local interactions modifying their coefficients. By either solving the Landau kinetic equation or performing a random phase approximation (RPA) analysis of the density-density correlator, the Matsubara density response is given by~\cite{GV2008Book}
\begin{align}
\Pi_{\tau\tau}^{\textrm{RPA}}(\mathtt{i}\omega,\boldsymbol{k})=\frac{\Pi_{\tau\tau}(\mathtt{i}\omega,\boldsymbol{k})}{1-(F_{0}/\mathscr{D}_{F})\Pi_{\tau\tau}(\mathtt{i}\omega,\boldsymbol{k})},
\label{eq:_FL_RPA}
\end{align}
where $F_{0}$ is the (dimensionless) Landau parameter in the $s$-wave channel, and $\mathscr{D}_{F}=\frac{\Omega_{d-1}}{(2\pi)^{d}}mk_{F}^{d-2}$ is the density of states at the Fermi surface. 

Introducing the dimensionless variables $\tilde{\omega}=\frac{m\omega}{k_{F}|\boldsymbol{k}|}$ and $\tilde{k}=\frac{|\boldsymbol{k}|}{2k_{F}}$, the Lindhard function takes the form~\cite{GV2008Book}
\begin{align}
&\Pi_{\tau\tau}(\mathtt{i}\omega,\boldsymbol{k})=\mathscr{D}_{F}\frac{\varPhi_{d}(\mathtt{i}\tilde{\omega}-\tilde{k})-\varPhi_{d}(\mathtt{i}\tilde{\omega}+\tilde{k})}{2\tilde{k}}\nonumber\\\textrm{with}\quad&\varPhi_{d}(z)=\begin{cases}
\displaystyle
\int_{0}^{1}\textrm{d}\lambda\lambda^{d-1}\frac{1}{2}\left(\frac{1}{z-\lambda}+\frac{1}{z+\lambda}\right) & d=1\\\\
\displaystyle
\int_{0}^{1}\textrm{d}\lambda\lambda^{d-1}\int_{0}^{\pi}\textrm{d}\theta\frac{\Omega_{d-2}}{\Omega_{d-1}}\frac{(\sin\theta)^{d-2}}{z-\lambda\cos\theta} & d\geq2
\end{cases}.
\end{align}
For $|\boldsymbol{k}|<2k_{F}$ (i.e., $\tilde{k}<1$), the Lindhard function admits the Taylor series
\begin{align}
\Pi_{\tau\tau}(\mathtt{i}\omega,\boldsymbol{k})=-\mathscr{D}_{F}\sum_{\ell=0}^{+\infty}\frac{\varPhi_{d}^{(2\ell+1)}(\mathtt{i}\tilde{\omega})}{(2\ell+1)!}\tilde{k}^{2\ell},
\end{align}
where $\Phi_{d}^{(n)}$ denotes the $n$-th derivative of $\Phi_{d}$. Substituting into Eq.~\eqref{eq:_FL_RPA}, the full static structure factor (normalized by the charge density $\langle\rho\rangle=\frac{\Omega_{d-1}}{(2\pi)^{d}}\frac{k_{F}^{d}}{d}$) acquires the long-wavelength expansion 
\begin{align}
\frac{S(\boldsymbol{k})}{\langle\rho\rangle}=-2d\tilde{k}\int_{-\infty}^{+\infty}\frac{\textrm{d}\tilde{\omega}}{2\pi}\frac{\Pi_{\tau\tau}(\mathtt{i}\omega,\boldsymbol{k})/\mathscr{D}_{F}}{1-F_{0}\Pi_{\tau\tau}(\mathtt{i}\omega,\boldsymbol{k})/\mathscr{D}_{F}}=\sum_{\ell=0}^{+\infty}\int_{-\infty}^{+\infty}\frac{\textrm{d}\tilde{\omega}}{2\pi}g_{2\ell+1}(\mathtt{i}\tilde{\omega})\tilde{k}^{2\ell+1},
\end{align}
where, for example,
\begin{align}
g_{1}(\mathtt{i}\tilde{\omega})=\frac{2d\varPhi_{d}^{(1)}(\mathtt{i}\tilde{\omega})}{1+F_{0}\varPhi_{d}^{(1)}(\mathtt{i}\tilde{\omega})},\qquad g_{3}(\mathtt{i}\tilde{\omega})=\frac{2d}{(1+F_{0}\varPhi_{d}^{(1)}(\mathtt{i}\tilde{\omega}))^{2}}\frac{\varPhi_{d}^{(3)}(\mathtt{i}\tilde{\omega})}{3!},
\end{align}
and similarly for higher orders. This makes manifest that only odd powers of $|\boldsymbol{k}|$ appear in $S(\boldsymbol{k})$, in agreement with the structure found in the free-fermion case. 

\begin{figure}
    \centering
 \includegraphics[width=0.88\linewidth]{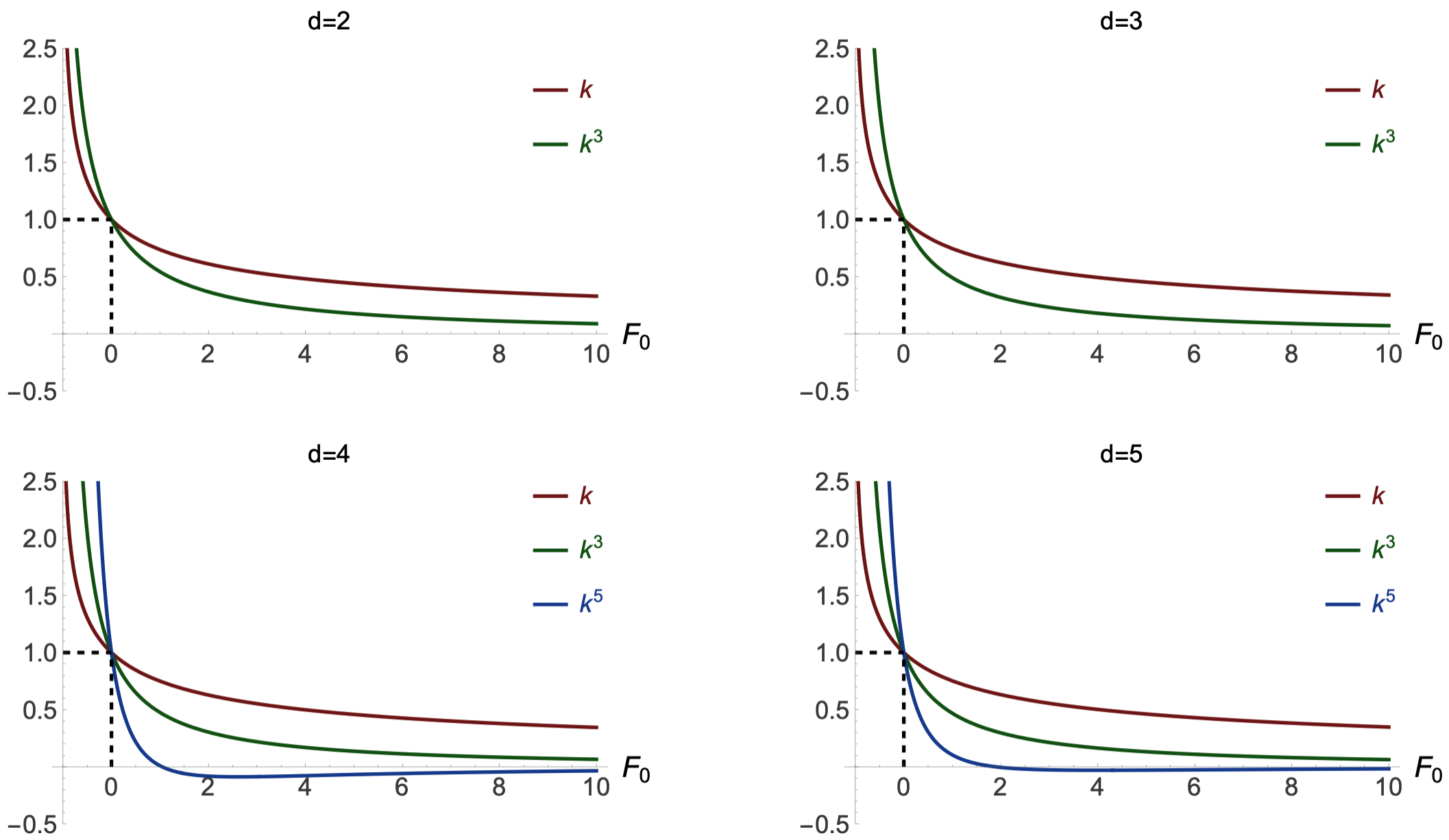}
    \caption{Coefficients in the long-wavelength expansion of the static structure factor $S(k)$ for Landau Fermi liquids in spatial dimensions $d=2,3,4,5$, shown as functions of the Landau parameter $F_{0}$. All coefficients are normalized by their free-Fermi-gas values (given in Eq.~\eqref{eq:_FG_SSF-II}), and therefore return to unity at $F_{0}=0$.}
   
    \label{fig:_FL_SSF}
\end{figure}

In $d=2$, the dependence of the linear $|\boldsymbol{k}|$ term on the Landau parameter $F_{0}$ has been analyzed in Refs.~\cite{FSEE_swingle2,Cai_FS_2024,Wu2025CFS}. Here, we provide a more systematic study of the leading few terms across various spatial dimensions. The results are shown in FIG.~\ref{fig:_FL_SSF}, where we restrict to $F_{0}>-1$ (recall that $F_{0}=-1$ marks the Pomeranchuk instability).

\section{Angle Function for Parallelepiped}

We explicitly derive the universal angle dependence of a trihedral corner, as stated in Eq.~\eqref{eq:_angle_fun_3d-I}, using the subtraction scheme introduced in Eq.~\eqref{eq:_corner_scheme_3d}. As explained in the main text, the key step is to evaluate Eq.~\eqref{eq:_SSF_pair_3d}. To separate the universal geometric dependence from the system-dependent coefficient in Eq.~\eqref{eq:_3th_mom}, we rewrite the integral in the form
\begin{align}
\Xi_{\textrm{AG}}=\int\textrm{d}^{3}\boldsymbol{r}S(\boldsymbol{r})\int_{\textrm{A}}\textrm{d}^{3}\boldsymbol{r}_{1}\int_{\textrm{G}}\textrm{d}^{3}\boldsymbol{r}_{2}\delta^{3}(\boldsymbol{r}_{1}-\boldsymbol{r}_{2}-\boldsymbol{r})=\mathcal{I}(\theta_{1,2},\theta_{2,3},\theta_{1,3})\int_{0}^{+\infty}\textrm{d}rr^{5}S(r).
\end{align}
Here, the geometric factor is defined by
\begin{align}
\mathcal{I}(\theta_{1,2},\theta_{2,3},\theta_{1,3})=\int\textrm{d}\Omega_{\hat{\boldsymbol{r}}}\int_{\textrm{A}}\textrm{d}^{3}\boldsymbol{r}_{1}\int_{\textrm{G}}\textrm{d}^{3}\boldsymbol{r}_{2}\delta^{3}(\boldsymbol{r}_{1}-\boldsymbol{r}_{2}-\hat{\boldsymbol{r}}),
\label{eq:_3d_corner_int_1}
\end{align}
where $\hat{\boldsymbol{r}}$ is a unit vector, and $\textrm{d}\Omega_{\hat{\boldsymbol{r}}}$ denotes the corresponding solid-angle measure.

The integral of $\delta^{3}(\boldsymbol{r}_{1}-\boldsymbol{r}_{2}-\hat{\boldsymbol{r}})$
over $\boldsymbol{r}_{1}\in\textrm{A}$ and $\boldsymbol{r}_{2}\in\textrm{G}$
admits a geometric interpretation: it computes the volume $V_{\textrm{A}\cap\textrm{G}}(\hat{\boldsymbol{r}})$
of the overlap between region A and region G after translating G by
the unit vector $\hat{\boldsymbol{r}}$. Clearly, this overlap is nonzero only when $\hat{\boldsymbol{r}}$ lies inside $\textrm{A}$.
To express $V_{\textrm{A}\cap\textrm{G}}(\hat{\boldsymbol{r}})$ conveniently, we decompose $\hat{\boldsymbol{r}}$
in the basis of $\hat{\boldsymbol{e}}_{1},\hat{\boldsymbol{e}}_{2},\hat{\boldsymbol{e}}_{3}$
: 
\begin{equation}
\hat{\boldsymbol{r}}=\sum_{i=1}^{3}u_{i}\hat{\boldsymbol{e}}_{i}.
\end{equation}
The coefficients $u_{i}\geq0$ can be written as $u_{i}=\hat{\boldsymbol{r}}\cdot\hat{\boldsymbol{\tau}}_{i}$,
where the reciprocal basis vectors are given by
\begin{equation}
\hat{\boldsymbol{\tau}}_{1}=\frac{\hat{\boldsymbol{e}}_{2}\times\hat{\boldsymbol{e}}_{3}}{\hat{\boldsymbol{e}}_{1}\cdot(\hat{\boldsymbol{e}}_{2}\times\hat{\boldsymbol{e}}_{3})},\quad\hat{\boldsymbol{\tau}}_{2}=\frac{\hat{\boldsymbol{e}}_{3}\times\hat{\boldsymbol{e}}_{1}}{\hat{\boldsymbol{e}}_{1}\cdot(\hat{\boldsymbol{e}}_{2}\times\hat{\boldsymbol{e}}_{3})},\quad\hat{\boldsymbol{\tau}}_{3}=\frac{\hat{\boldsymbol{e}}_{1}\times\hat{\boldsymbol{e}}_{2}}{\hat{\boldsymbol{e}}_{1}\cdot(\hat{\boldsymbol{e}}_{2}\times\hat{\boldsymbol{e}}_{3})},
\end{equation}
which satisfy the orthonormality condition $\hat{\boldsymbol{\tau}}_{i}\cdot\hat{\boldsymbol{e}}_{j}=\delta_{ij}$.
It is also convenient to introduce the Gram matrix
\begin{equation}
G_{ij}^{+++}=\hat{\boldsymbol{e}}_{i}\cdot\hat{\boldsymbol{e}}_{j}=\left(\begin{array}{ccc}
1 & \cos\theta_{1,2} & \cos\theta_{1,3}\\
\cos\theta_{1,2} & 1 & \cos\theta_{2,3}\\
\cos\theta_{1,3} & \cos\theta_{2,3} & 1
\end{array}\right)
\label{}
\end{equation}
as in Eq.~\eqref{eq:_Gram_matrix}. The volume of the unit cell spanned by $\hat{\boldsymbol{e}}_{i}$ is
\[
V_{\hat{\boldsymbol{e}}}=|\hat{\boldsymbol{e}}_{1}\cdot(\hat{\boldsymbol{e}}_{2}\times\hat{\boldsymbol{e}}_{3})|=\sqrt{\det(G^{+++})}.
\]
The overlap volume is then $V_{\textrm{A}\cap\textrm{G}}(\hat{\boldsymbol{r}})=(u_{1}u_{2}u_{3})V_{\hat{\boldsymbol{e}}}$,
allowing us to rewrite Eq.~\eqref{eq:_3d_corner_int_1} as
\begin{equation}
\mathcal{I}(\theta_{1,2},\theta_{2,3},\theta_{1,3})=\int_{\textrm{A}}\textrm{d}\Omega_{\hat{\boldsymbol{r}}}V_{\textrm{A}\cap\textrm{G}}(\hat{\boldsymbol{r}})=V_{\hat{\boldsymbol{e}}}\int_{\textrm{A}}\textrm{d}\Omega_{\hat{\boldsymbol{r}}}(\hat{\boldsymbol{r}}\cdot\hat{\boldsymbol{\tau}}_{1})(\hat{\boldsymbol{r}}\cdot\hat{\boldsymbol{\tau}}_{2})(\hat{\boldsymbol{r}}\cdot\hat{\boldsymbol{\tau}}_{3}).\label{eq:_3d_corner_int_2}
\end{equation}

The challenging aspect of the integral in Eq.~\eqref{eq:_3d_corner_int_2}
lies in handling the geometric constraint that $\hat{\boldsymbol{r}}$
lies within region A. To address this, we introduce barycentric coordinates
on the unit sphere:
\begin{equation}
\hat{\boldsymbol{r}}(v_{1},v_{2})=\frac{v_{1}\hat{\boldsymbol{e}}_{1}+v_{2}\hat{\boldsymbol{e}}_{2}+v_{3}\hat{\boldsymbol{e}}_{3}}{\rho(v_{1},v_{2})}\quad\textrm{with}\quad\rho(v_{1},v_{2})=|v_{1}\hat{\boldsymbol{e}}_{1}+v_{2}\hat{\boldsymbol{e}}_{2}+v_{3}\hat{\boldsymbol{e}}_{3}|,
\end{equation}
where $v_{3}=1-v_{1}-v_{2}$, and $v_{i}\geq0$ for $i=1,2,3$. The
solid-angle element in $(v_{1},v_{2})$ coordinates is given by
\begin{equation}
\textrm{d}\Omega_{\hat{\boldsymbol{r}}}=\left|\frac{\partial\hat{\boldsymbol{r}}}{\partial v_{1}}\times\frac{\partial\hat{\boldsymbol{r}}}{\partial v_{2}}\right|\textrm{d}v_{1}\textrm{d}v_{2}=\frac{V_{\hat{\boldsymbol{e}}}}{\rho^{3}}\textrm{d}v_{1}\textrm{d}v_{2}.
\end{equation}
In terms of the barycentric coordinates, Eq.~\eqref{eq:_3d_corner_int_2}
becomes 
\begin{equation}
\mathcal{I}(\theta_{1,2},\theta_{2,3},\theta_{1,3})=V_{\hat{\boldsymbol{e}}}^{2}\int_{\Delta_{2}}\textrm{d}v_{1}\textrm{d}v_{2}\frac{v_{1}v_{2}v_{3}}{\rho^{6}}=I_{3}(G^{+++})
\end{equation}
where $I_{3}(G)$ is the functional defined in Eq.~\eqref{eq:_Int_Gram_3d}.

A similar analysis applies to $\Xi_{\textrm{BH}}$, $\Xi_{\textrm{CE}}$, and $\Xi_{\textrm{DF}}$. Collecting all contributions, we obtain the universal angle function Eq.~\eqref{eq:_angle_fun_3d-I} for an arbitrary parallelepiped
\begin{flalign}
f_{3}(\theta_{1,2},\theta_{2,3},\theta_{1,3})=2\left[\begin{array}{c}
\mathcal{I}(\theta_{1,2},\theta_{2,3},\theta_{1,3})+\mathcal{I}(\theta_{1,2},\pi-\theta_{2,3},\pi-\theta_{1,3})\\
+\mathcal{I}(\pi-\theta_{1,2},\theta_{2,3},\pi-\theta_{1,3})+\mathcal{I}(\pi-\theta_{1,2},\pi-\theta_{2,3},\theta_{1,3})
\end{array}\right].
\end{flalign}


\section{Angle Function for  Parallelotope}

We introduce a method, complementary to the subtraction scheme in Fig.~\ref{fig:_corner}, to derive the total corner charge fluctuations associated with an arbitrary parallelotope in general dimensions. 

In general, the bipartite fluctuations associated with a subsystem $\Sigma$ can be written as
\begin{equation}
\mathcal{F}_{\Sigma}=\int_{\Sigma}\textrm{d}^{d}\boldsymbol{r}_{1}\int_{\Sigma}\textrm{d}^{d}\boldsymbol{r}_{2}S(\boldsymbol{r}_{1}-\boldsymbol{r}_{2})=\int_{\Sigma}\textrm{d}^{d}\boldsymbol{r}_{1}\int_{\Sigma}\textrm{d}^{d}\boldsymbol{r}_{2}\int\frac{\textrm{d}^{d}\boldsymbol{k}}{(2\pi)^{d}}e^{-\mathtt{i}\boldsymbol{k}\cdot(\boldsymbol{r}_{1}-\boldsymbol{r}_{2})}S(\boldsymbol{k})=\int\frac{\textrm{d}^{d}\boldsymbol{k}}{(2\pi)^{d}}S(\boldsymbol{k})|\Theta_{\Sigma}(\boldsymbol{k})|^{2},\label{eq:_BF_bulk}
\end{equation}
where $\Theta_{\Sigma}(\boldsymbol{k})$ denotes the Fourier transform
of the characteristic function of $\Sigma$, defined by 
\begin{equation}
\Theta_{\Sigma}(\boldsymbol{k})=\int_{\Sigma}\textrm{d}^{d}\boldsymbol{r}e^{\mathtt{i}\boldsymbol{r}\cdot\boldsymbol{k}}.
\end{equation}

Any $d$-dimensional parallelotope can be specified by a set of linearly independent unit vectors $\hat{\boldsymbol{e}}_{1},\ldots,\hat{\boldsymbol{e}}_{d}$,
each corresponding to the direction of one of its edges. We further introduce the dual basis vectors $\hat{\boldsymbol{\tau}}_{i}$,
defined through the orthonormality condition $\hat{\boldsymbol{e}}_{i}\cdot\hat{\boldsymbol{\tau}}_{j}=\delta_{ij}$.
In this coordinate system, real-space and momentum-space vectors are 
\begin{equation}
\boldsymbol{r}=\sum_{i=1}^{d}r_{i}\hat{\boldsymbol{e}}_{i},\qquad\boldsymbol{k}=\sum_{i=1}^{d}k_{i}\hat{\boldsymbol{\tau}}_{i},
\end{equation}
with $r_{i}=\boldsymbol{r}\cdot\hat{\boldsymbol{\tau}}_{i}$
and $k_{i}=\boldsymbol{k}\cdot\hat{\boldsymbol{e}}_{i}$. The characteristic function of the parallelotope then takes the form 
\begin{equation}
\Theta_{\Sigma}(\boldsymbol{k})=V_{\hat{\boldsymbol{e}}}\prod_{i=1}^{d}\int_{0}^{L}\textrm{d}r_{i}e^{\mathtt{i}r_{i}k_{i}}=V_{\hat{\boldsymbol{e}}}\prod_{i=1}^{d}\frac{\mathtt{i}}{k_{i}}(1-e^{\mathtt{i}k_{i}L}),
\end{equation}
where $V_{\hat{\boldsymbol{e}}}=|\det(\hat{\boldsymbol{e}}_{j}^{i})|$ is the volume spanned by the unit vectors $\hat{\boldsymbol{e}}_{1},\ldots,\hat{\boldsymbol{e}}_{d}$.

We can then use the following identity to rewrite the factor
$|\Theta_{\Sigma}(\boldsymbol{k})|^{2}$ appearing in
Eq.~\eqref{eq:_BF_bulk}:
\begin{equation}
|\Theta_{\Sigma}(\boldsymbol{k})|^{2}=V_{\hat{\boldsymbol{e}}}^{2}\prod_{i=1}^{d}\frac{2-2\cos(k_{i}L)}{k_{i}^{2}}=V_{\hat{\boldsymbol{e}}}^{2}\prod_{i=1}^{d}\int_{0}^{L}\textrm{d}u_{i}2(L-u_{i})\cos(k_{i}u_{i})=V_{\hat{\boldsymbol{e}}}^{2}\prod_{i=1}^{d}\int_{0}^{L}\textrm{d}u_{i}(L-u_{i})\sum_{s_{i}=\pm1}e^{\mathtt{i}s_{i}k_{i}u_{i}}.
\end{equation}
Using this representation and the Fourier transform of $S(\boldsymbol{k})$, we can rewrite the bipartite fluctuations in
real space as
\begin{flalign}
\mathcal{F}_{\Sigma} & =\sum_{s_{1},\ldots,s_{d}=\pm}V_{\hat{\boldsymbol{e}}}^{2}\left(\prod_{i=1}^{d}\int_{0}^{L}\textrm{d}u_{i}(L-u_{i})\right)S\left(\sum_{i=1}^{d}s_{i}u_{i}\hat{\boldsymbol{e}}_{i}\right)\nonumber \\
 & =\sum_{s_{1},\ldots,s_{d}=\pm}V_{\hat{\boldsymbol{e}}}^{2}\left(\prod_{i=1}^{d}\int_{0}^{L}\textrm{d}u_{i}(L-u_{i})\right)\int\textrm{d}r\delta\left(r-\left|\sum_{i=1}^{d}s_{i}u_{i}\hat{\boldsymbol{e}}_{i}\right|\right)S(r).
\end{flalign}
where the $S(\boldsymbol{r})$ is assumed to be rotationally invariant.

Since our interest lies in the dimensionless contribution, we discard all power-law terms in $L$, which yields
\begin{equation}
\mathcal{F}_{\Sigma}\supset(-1)^{d}\sum_{s_{1},\ldots,s_{d}=\pm}V_{\hat{\boldsymbol{e}}}^{2}\left(\prod_{i=1}^{d}\int_{0}^{L}\textrm{d}u_{i}u_{i}\right)\int\textrm{d}r\delta\left(r-\left|\sum_{i=1}^{d}s_{i}u_{i}\hat{\boldsymbol{e}}_{i}\right|\right)S(r).
\end{equation}
To isolate the universal angle dependence from the
system-dependent coefficient,
we perform a radial-simplex
decomposition of the integration variables.
Specifically, we change variables from $u_{1},u_{2},\ldots,u_{d}$ to a radial variable $\lambda$
and barycentric coordinates $v_{1},\ldots,v_{d-1}$ by setting $u_{i}=\lambda v_{i}$.
With this substitution, the integration measure becomes
\begin{equation}
\prod_{i=1}^{d}\int_{0}^{L}\textrm{d}u_{i}=\left(\prod_{i=1}^{d}\int_{0}^{1}\textrm{d}v_{i}\right)\delta\left(\sum_{i=1}^{d}v_{i}-1\right)\int_{0}^{\textrm{min}(L/v_{1},\ldots,L/v_{d})}\textrm{d}\lambda\lambda^{d-1}.
\end{equation}
Carrying out the $\lambda$ integration yields
\begin{equation}
\int\textrm{d}\lambda\lambda^{2d-1}\delta\left(r-\lambda\left|\sum_{i=1}^{d}s_{i}v_{i}\hat{\boldsymbol{e}}_{i}\right|\right)=\left|\sum_{i=1}^{d}s_{i}v_{i}\hat{\boldsymbol{e}}_{i}\right|^{-2d}r^{2d-1}.
\end{equation}
As a result, the universal dimensionless contribution takes the form
\begin{equation}
\mathcal{F}_{\Sigma}\supset(-1)^{d}\sum_{s_{1},\ldots,s_{d}=\pm}V_{\hat{\boldsymbol{e}}}^{2}\left(\prod_{i=1}^{d}\int_{0}^{1}\textrm{d}v_{i}\right)\delta\left(\sum_{i=1}^{d}v_{i}-1\right)\left|\sum_{i=1}^{d}s_{i}v_{i}\hat{\boldsymbol{e}}_{i}\right|^{-2d}\int\textrm{d}rr^{2d-1}S(r).
\end{equation}

At this stage, it is convenient to introduce the Gram matrices $G_{ij}^{\{\mathbf{s}\}}$ defined in Eq.~\eqref{eq:_Gram_matrix}. By construction, one has $G_{ij}^{\{\mathbf{s}\}}=G_{ij}^{\{-\mathbf{s}\}}$, and for all sign choices $\det G_{ij}^{\{\mathbf{s}\}}=V_{\hat{\boldsymbol{e}}}^{2}=|\det(\hat{\boldsymbol{e}}_{j}^{i})|^{2}$. With these definitions, the final result can be written compactly as $\mathcal{F}_{\Sigma}\supset f_{d}(\theta_{i,j})\varUpsilon_{d}$, where $\varUpsilon_{d}$ is the $d$-th radial moment defined in Eq.~\eqref{eq:_dth_mom}, and the universal angle function $f_{d}(\theta_{i,j})$ is given in Eq.~\eqref{eq:_angle_fun_dd-I}.

\section{Quantum Metric from Dihedral Wedge}

\subsection{General considerations}
In this section, we illustrate the protocol for extracting the coefficient $\mathsf{S}_{2}$ using the wedge configuration shown in FIG.~\ref{fig:_wedge}. (a). We consider a finite dihedral wedge $\textrm{A}=\textrm{A}^{\prime}\times[0,L_{w}]^{d-2}$ in $d$ spatial dimensions, where $\textrm{A}^{\prime}$  is a two-dimensional subregion resembling the one in  FIG.~\ref{fig:_corner}(a). We define the combination
\begin{flalign}
\mathcal{W}_{d}=\mathcal{F}_{\textrm{A}}+\mathcal{F}_{\textrm{B}}+\mathcal{F}_{\textrm{C}}+\mathcal{F}_{\textrm{D}}-\mathcal{F}_{\textrm{AB}}-\mathcal{F}_{\textrm{CD}}-\mathcal{F}_{\textrm{BC}}-\mathcal{F}_{\textrm{AD}}+\mathcal{F}_{\textrm{ABCD}},
\label{eq:_wedge_scheme}
\end{flalign}
which becomes $\mathcal{C}_2$ in $d=2$. From the conservation of charge in the total system, we have $\mathcal{F}_{\rm A} = -\Xi_{{\rm A}\bar{\rm A}}$, where $\Xi_{{\rm A}\bar{\rm A}}$ is the charge correlation between region $\rm A$ and its complement $\bar{\rm A}$ (see also the definition in Eq.~\eqref{eq:_SSF_pair_3d} for $d=3$). One can then simplify the wedge term $\mathcal{W}_d$ as
\begin{align}
    \frac{\mathcal{W}_{d}}{2}=\Xi_{\rm AC}+\Xi_{\rm BD}=\int_{\textrm{A}}\textrm{d}^{d}\boldsymbol{r}_{1}\int_{\textrm{C}}\textrm{d}^{d}\boldsymbol{r}_{2}S(\boldsymbol{r}_{1}-\boldsymbol{r}_{2})+\int_{\textrm{B}}\textrm{d}^{d}\boldsymbol{r}_{1}\int_{\textrm{D}}\textrm{d}^{d}\boldsymbol{r}_{2}S(\boldsymbol{r}_{1}-\boldsymbol{r}_{2}),
\end{align}
where $\rm A$ (resp. $\rm B$) and $\rm C$ (resp. $\rm D$) are two wedge sharing regions.
We first analyze the first term:
\begin{align}
\int_{\textrm{A}}\textrm{d}^{d}\boldsymbol{r}_{1}\int_{\textrm{C}}\textrm{d}^{d}\boldsymbol{r}_{2}S(\boldsymbol{r}_{1}-\boldsymbol{r}_{2})=\int_{\textrm{A}^{\prime}}\textrm{d}^{2}\boldsymbol{r}_{1}^{\prime}\int_{\textrm{C}^{\prime}}\textrm{d}^{2}\boldsymbol{r}_{2}^{\prime}\left(\prod_{i=1}^{d-2}\int_{0}^{L_{w}}\textrm{d}z_{1}^{i}\int_{0}^{L_{w}}\textrm{d}z_{2}^{i}\right)S(\boldsymbol{r}_{1}^{\prime}-\boldsymbol{r}_{2}^{\prime},\boldsymbol{z}_{1}-\boldsymbol{z}_{2}).
\end{align}
Changing variables to $\boldsymbol{a}=\boldsymbol{z}_{1}-\boldsymbol{z}_{2}$ and $\boldsymbol{b}=\boldsymbol{z}_{1}+\boldsymbol{z}_{2}$ yields
\begin{align}
\int_{\textrm{A}}\textrm{d}^{d}\boldsymbol{r}_{1}\int_{\textrm{C}}\textrm{d}^{d}\boldsymbol{r}_{2}S(\boldsymbol{r}_{1}-\boldsymbol{r}_{2})=\int_{\textrm{A}^{\prime}}\textrm{d}^{2}\boldsymbol{r}_{1}^{\prime}\int_{\textrm{C}^{\prime}}\textrm{d}^{2}\boldsymbol{r}_{2}^{\prime}\left(\prod_{i=1}^{d-2}\int_{-L_{w}}^{L_{w}}\textrm{d}a^{i}(L_{w}-|a^{i}|)\right)S(\boldsymbol{r}_{1}^{\prime}-\boldsymbol{r}_{2}^{\prime},\boldsymbol{a}).
\end{align}
The results in the main text follow by taking the limit
\begin{align}
\lim_{L_{w}\rightarrow\infty}\frac{1}{L_{w}^{d-2}}\int_{\textrm{A}}\textrm{d}^{d}\boldsymbol{r}_{1}\int_{\textrm{C}}\textrm{d}^{d}\boldsymbol{r}_{2}S(\boldsymbol{r}_{1}-\boldsymbol{r}_{2})=\int_{\textrm{A}^{\prime}}\textrm{d}^{2}\boldsymbol{r}_{1}^{\prime}\int_{\textrm{C}^{\prime}}\textrm{d}^{2}\boldsymbol{r}_{2}^{\prime}\int_{\mathbb{R}^{d-2}}\textrm{d}^{d-2}\boldsymbol{a}S(\boldsymbol{r}_{1}^{\prime}-\boldsymbol{r}_{2}^{\prime},\boldsymbol{a}).
\end{align}
An important simplification underlying the above formula is the observation that if a function $g(\boldsymbol{r})$ decays as a power law $g(\boldsymbol{r})\sim1/|\boldsymbol{r}|^{\alpha}$ with $\alpha>1$, then
\begin{align}
\lim_{L_{w}\rightarrow\infty}\int_{-L_{w}}^{L_{w}}\textrm{d}a\frac{|a|}{L_{w}}g(\boldsymbol{x},a)\sim\lim_{L_{w}\rightarrow\infty}\int_{-L_{w}}^{L_{w}}\textrm{d}a\frac{|a|}{L_{w}}\frac{1}{(|\boldsymbol{x}|^{2}+a^{2})^{\frac{\alpha}{2}}}=\lim_{L_{w}\rightarrow\infty}\frac{(L_{w}^{2}+|\boldsymbol{x}|^{2})^{1-\frac{\alpha}{2}}-|\boldsymbol{x}|^{2-\alpha}}{(1-\frac{\alpha}{2})L_{w}}=0,
\end{align}
and the same conclusion holds when $g(\boldsymbol{r})$ decays exponentially. Here we have made this assumption about the decay of $S(\boldsymbol{r})$, which should be valid for most gapless and (surely) gapped systems. Notice that the same analysis applies to $\Xi_{\rm BD}$, so altogether we obtain
\begin{equation}\label{supp_eq: wedge general without isotropy}
    \lim_{L_w\rightarrow\infty}\frac{\mathcal{W}_d}{2 L_w^{d-2}} =\left( \int_{\textrm{A}^{\prime}}\textrm{d}^{2}\boldsymbol{r}_{1}^{\prime}\int_{\textrm{C}^{\prime}}\textrm{d}^{2}\boldsymbol{r}_{2}^{\prime}+ \int_{\textrm{B}^{\prime}}\textrm{d}^{2}\boldsymbol{r}_{1}^{\prime}\int_{\textrm{D}^{\prime}}\textrm{d}^{2}\boldsymbol{r}_{2}^{\prime}\right)\mathfrak{S}(\boldsymbol{r}_{1}^{\prime}-\boldsymbol{r}_{2}^{\prime}),\quad\text{with}\quad\mathfrak{S}(\boldsymbol{r})\equiv \int_{\mathbb{R}^{d-2}}\textrm{d}^{d-2}\boldsymbol{a}S(\boldsymbol{r},\boldsymbol{a}),
\end{equation}
which reduces the problem effectively to that about corner charge fluctuations in two dimensions.  Assuming isotropy, $\mathfrak{S}(\boldsymbol{r}) = \mathfrak{S}(r)$ is a function of $r\equiv |\boldsymbol{r}|$ alone, hence
\begin{equation}
\begin{split}
    \int_{\textrm{A}^{\prime}}\textrm{d}^{2}\boldsymbol{r}_{1}^{\prime}\int_{\textrm{C}^{\prime}}\textrm{d}^{2}\boldsymbol{r}_{2}^{\prime} \mathfrak{S}(\boldsymbol{r}_{1}^{\prime}-\boldsymbol{r}_{2}^{\prime}) &= \int_0^\infty \textrm{d}r \mathfrak{S}(r) \Big[ \int_{\rm A'} \textrm{d}^{2}\boldsymbol{r}_1 \int_{\rm C'}\textrm{d}^{2}\boldsymbol{r}_2 \delta(|\boldsymbol{r}_1-\boldsymbol{r}_2|-r) \Big] \\
&= \int_0^\infty \textrm{d}r r^3\mathfrak{S}(r) \Big[ \int_{\rm A'} \textrm{d}^{2}\boldsymbol{r}_1 \int_{\rm C'}\textrm{d}^{2}\boldsymbol{r}_2 \delta(|\boldsymbol{r}_1-\boldsymbol{r}_2|-1) \Big] \\
&= \frac{1}{2}\mathtt{f}(\pi-\theta)\int_0^\infty \textrm{d}r r^3\mathfrak{S}(r)
\end{split}
\end{equation}
where $\theta$ is the wedge angle as depicted in FIG. \ref{fig:_wedge}(a),  and the angle function $\mathtt{f}(\pi-\theta) = 1-\theta\cot\theta$ is obtained from doing the integral inside the square bracket, whose derivation can be found in Ref. \cite{estienne2022cornering}. Altogether we have
\begin{equation}\label{supp_eq: wedge universal angle dependence}
    \lim_{L_w\rightarrow\infty}\frac{\mathcal{W}_d}{L_w^{d-2}} = -f_2(\theta)\int_0^\infty \textrm{d}r r^3\mathfrak{S}(r) = \frac{2}{\pi}\mathsf{S}_2 f_2(\theta)
\end{equation}
where $\mathsf{S}_{2}$ is the coefficient of the $|\boldsymbol{k}|^{2}$ term in the static structure factor $S(\boldsymbol{k})$, and the universal angle function $f_{2}(\theta)$ is given in Eq.~\eqref{eq:_angle_fun_2d}. This is Eq.~\eqref{eq:_QM_wedge-I} in the main text. The above result thus generalizes the universal angle dependence of corner charge fluctuations in two dimensions to arbitrary dimensions in terms of the wedge-corner charge fluctuations. 

\subsection{Anisotropic systems and lattice partition scheme}
Notice that Eq.~\eqref{supp_eq: wedge universal angle dependence} is derived under the assumption of full isotropy. For generic lattice models, including band insulators, modifications must be taken into account, and the small-wedge-angle limit becomes particularly important. Let us return to Eq.~\eqref{supp_eq: wedge general without isotropy} and take the wedge angle $\theta \rightarrow 0$, in which case the first term involving the correlation between $\rm A'$ and $\rm C'$ can be neglected when compared to the second term because regions $\rm A'$ and $C'$ are diminishing in size. Focusing on the second term in Eq.~\eqref{supp_eq: wedge general without isotropy}, 
\begin{equation}
\begin{split}
    \int_{\textrm{B}^{\prime}}\textrm{d}^{2}\boldsymbol{r}_{1}^{\prime}\int_{\textrm{D}^{\prime}}\textrm{d}^{2}\boldsymbol{r}_{2}^{\prime} \mathfrak{S}(\boldsymbol{r}_{1}^{\prime}-\boldsymbol{r}_{2}^{\prime}) =  \int'{\rm d}^2  \boldsymbol{r} \mathfrak{S}(\boldsymbol{r})\int'{\rm d}^2 \boldsymbol{R}\overset{\theta\rightarrow 0}{=} \frac{1}{2\theta}\int{\rm d}^2\boldsymbol{r} r^2_{\perp} \mathfrak{S}(\boldsymbol{r}).
\end{split}
\end{equation}
In the first equality, we have changed variables to $\boldsymbol{R} = (\boldsymbol{r}'_1+\boldsymbol{r}'_2)/2$ and $\boldsymbol{r}=\boldsymbol{r}'_1-\boldsymbol{r}'_2$, and the symbol $\int'$ indicates that the integral is constrained by requiring $\boldsymbol{r}'_1\in{\rm B'}$ and $\boldsymbol{r}'_2\in{\rm D'}$. For any fixed $\boldsymbol{r}$, it is straightforward to see that the $\boldsymbol{R}$-integral evaluates to $r^2_\perp\cot\theta-r_\perp r_{\parallel}$, where $r_\perp$($r_\parallel$) is the component of $\boldsymbol{r}$ in the direction perpendicular (parallel) to the partition line separating regions $\rm A'B'$ and $\rm C'D'$. Finally, in the limit $\theta\rightarrow 0$, we can further extend the $\boldsymbol{r}$-integral over the entire two-dimensional plane, leading to the second equality (where only the leading divergent term is kept). Notice that, for $\mathsf{S}^{ij}_2 k_ik_j\subset S(\boldsymbol{k})$, we have $\mathsf{S}^{\perp\perp}_2 = -\frac{1}{2}\int{\rm d}^2\boldsymbol{r} r^2_{\perp} \mathfrak{S}(\boldsymbol{r})$. Hence
\begin{equation}
    \mathsf{S}^{\perp\perp}_2 =\lim_{\theta\rightarrow0}\lim_{L_w\rightarrow\infty} \frac{-\theta\mathcal{W}_d}{2L_w^{d-2}},
\end{equation}
which is Eq.~\eqref{eq:_QM_wedge-III} in the main text.

While we have assumed continuous translation symmetry in the above derivation, it is easy to adapt the above reasoning on to a lattice system. Instead of performing continuous integrations over space, the calculation of charge fluctuations would involve lattice summations. We refer interested readers to the counting argument discussed around Eq. (14) of Ref. \cite{Tam2024Corner} for a complete lattice-level derivation. While the discussions in Refs. \cite{Tam2024Corner, Wu2025Corner} apply specifically to $d=2$ corner charge fluctuations, our explanation in this section should have made it clear that the wedge-corner charge fluctuations in general dimensions behave similarly. 

\begin{figure}
    \centering
    \includegraphics[width=\linewidth]{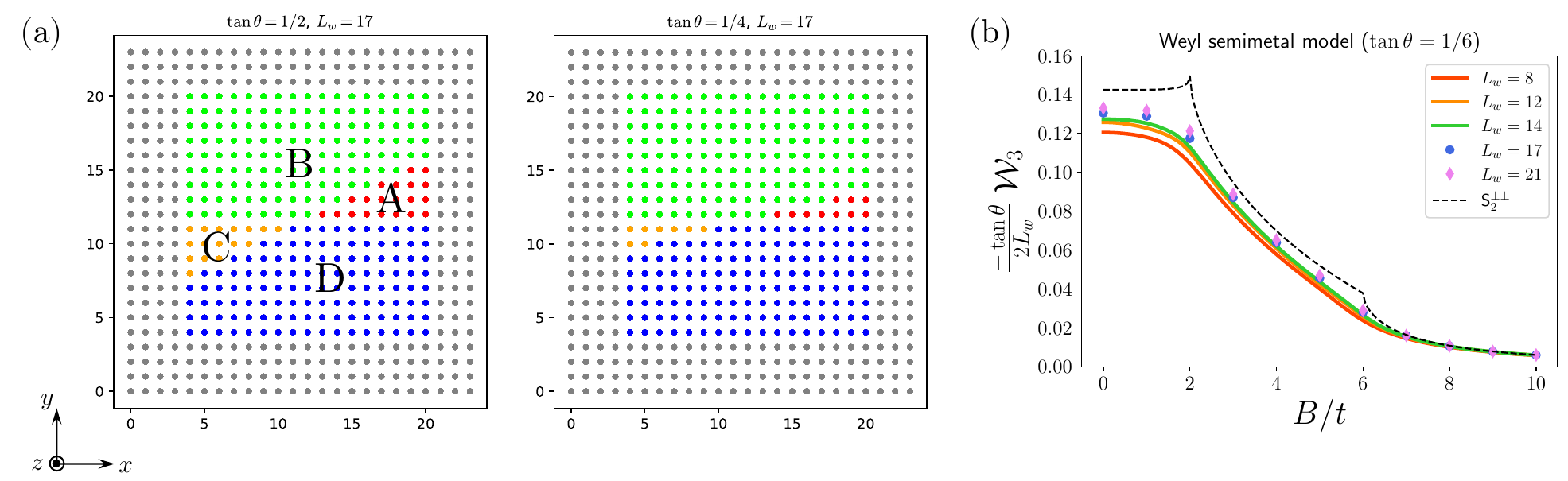}
    \caption{(a) Lattice partitions for the numerical calculation of $\mathcal{W}_3$ in the Weyl semimetal model discussed in the main text, see also FIG. \ref{fig:_wedge}(b). The wedge runs along the $z$-direction. (b) For a small wedge angle $\tan\theta=1/6$, and various wedge lengths $L_w$, $\mathcal{W}_3$ (normalized by $-\tan\theta/2L_w$) is calculated and compared with the integrated quantum metric $\mathsf{S}^{\bot\bot}_2$.}
    \label{supp_fig:Weyl}
\end{figure}

In FIG. \ref{supp_fig:Weyl}. (a), we show the precise lattice partition scheme we used for the numerical calculation of $\mathcal{W}_3$ in the Weyl semimetal model discussed in the main text. The lattice partition is specified by two partition planes cutting through the middle of lattice bonds. This choice makes the counting argument introduced in Ref. \cite{Tam2024Corner} exact, and thus best extracts the quantum metric in a lattice system. In FIG. \ref{supp_fig:Weyl}. (b), we further show results for an even smaller wedge angle than those featured in FIG. \ref{fig:_wedge}. (b) in the main text, with $\tan\theta=1/6$, and for a variety of system sizes $(L)$ and wedge lengths $(L_w=\lfloor 2L/3 \rfloor+1)$. Notice that the choice of a smaller wedge angle implies stronger finite-size effects in practice, parituclarly for the gapless phase when $B/t<6$.

\subsection{Remarks}
It is worth noting that $\mathcal{W}_{d}$ also contains a universal logarithmic subleading term in CFTs, although this contribution vanishes when taking the large-$L_{w}$ limit of $\mathcal{W}_{d}/L_{w}^{d-2}$ in $d>2$. To see this, let us examine
\begin{align}
\int_{\textrm{B}}\textrm{d}^{d}\boldsymbol{r}_{1}\int_{\textrm{D}}\textrm{d}^{d}\boldsymbol{r}_{2}S(\boldsymbol{r}_{1}-\boldsymbol{r}_{2})&\supset\int_{\textrm{B}^{\prime}}\textrm{d}^{2}\boldsymbol{r}_{1}^{\prime}\int_{\textrm{D}^{\prime}}\textrm{d}^{2}\boldsymbol{r}_{2}^{\prime}\left(\prod_{i=1}^{d-2}\int_{-L}^{L}\textrm{d}a^{i}(-|a^{i}|)\right)\frac{-C_{J}}{(|\boldsymbol{r}_{1}^{\prime}-\boldsymbol{r}_{2}^{\prime}|^{2}+|\boldsymbol{a}|^{2})^{2d}}\nonumber\\&=\int_{\textrm{B}^{\prime}}\textrm{d}^{2}\boldsymbol{r}_{1}^{\prime}\int_{\textrm{D}^{\prime}}\textrm{d}^{2}\boldsymbol{r}_{2}^{\prime}\frac{(-1)^{d-1}C_{J}}{(d-1)!}\frac{1}{|\boldsymbol{r}_{1}^{\prime}-\boldsymbol{r}_{2}^{\prime}|^{4}}.
\end{align}
The resulting expression can be viewed as the effective static structure factor of a $(2+1)$-dimensional system. Following the standard analysis for $(2+1)$-dimensional CFTs~\cite{estienne2022cornering,Wu2025Corner}, one finds
\begin{align}
    \mathcal{W}_{d}\supset C_{J}f_{d}(\theta,\pi/2,\ldots,\pi/2)\log(L),
\end{align}
where the angle function $f_{d}(\theta,\pi/2,\ldots,\pi/2)$ is given in Eq.~\eqref{eq:_angle_fun_dd-II}.

\section{Disorder Operator in the $\textrm{O}(3)$ Models} 



Although the focus of this work is on the bipartite fluctuations in Eq.~\eqref{eq:_2nd_cumulant}, corresponding to the small-$\chi$ limit of the disorder operator $\mathcal{U}_{\Sigma}(\chi)$ defined in Eq.~\eqref{eq:_dis_op}, where theoretical predictions can be derived analytically, it is also interesting to examine the behavior of $\mathcal{U}_{\Sigma}(\chi)$ over a broader range of $\chi$ numerically.

\begin{figure}
    \centering
    \includegraphics[width=0.8\linewidth]{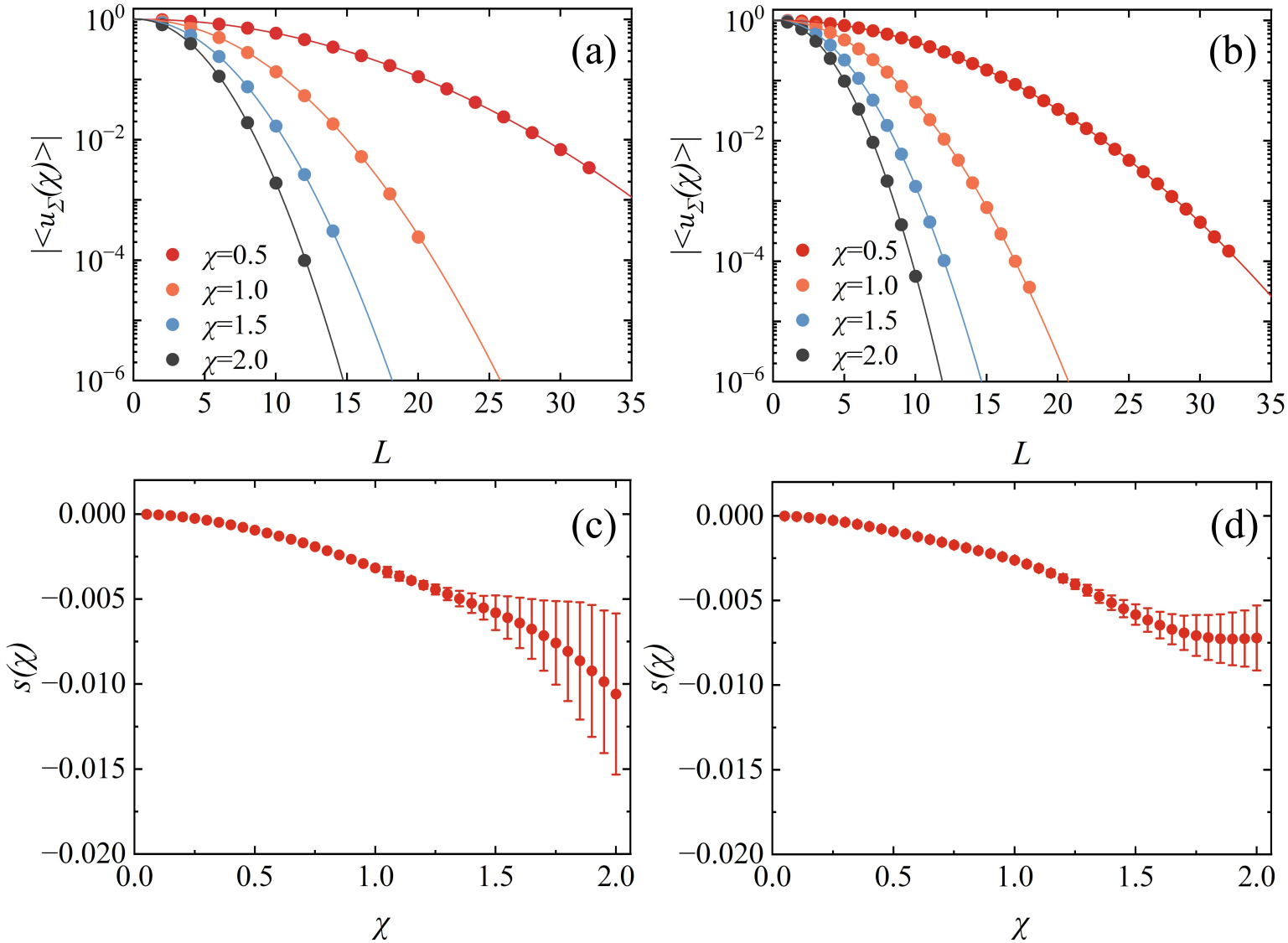}
    \caption{Disorder operator $\ensuremath{|\langle\mathcal{U}_{\Sigma}(\chi)\rangle|}$  versus subsystem size $L$ at the quantum critical point for $\chi=0.5, 1.0, 1.5, 2.0$ for (a) the columnar-dimerized (CD) and (b) double-cubic (DC) models. The data confirm that Eq.~\eqref{eq:disorder} holds for large $\chi$. Panels (c) and (d) show the logarithmic coefficient $s(\chi)$ extracted from the CD and DC models, respectively.}
    \label{fig:3D_CD_DC_operator}
\end{figure}

In the Heisenberg models, we evaluate the disorder operator $\mathcal{U}_{\Sigma}(\chi)=\prod_{j\in \Sigma}e^{\mathtt{i}\chi n_{j}}$ for the same parallelepiped geometry used in the main text ($\theta_{1,2}=\pi/4$ and $\theta_{1,3}=\theta_{2,3}=\pi/2$). While the small-$\chi$ regime is analyzed in the main text,
here we examine the behavior at larger $\chi$.
The quantum Monte Carlo results shown in FIG.~\ref{fig:3D_CD_DC_operator} demonstrate
that the scaling form Eq.~\eqref{eq:disorder}
remains valid over a wide range of $\chi$. The structure of Eq.~\eqref{eq:disorder} can be naturally interpreted in terms of the geometry of the subregion. The quadratic, linear, and logarithmic terms are expected to originate from surface, edge, and trihedral-corner contributions, respectively. Increasing $\chi$ then only modifies the coefficient functions $a(\chi)$, $b(\chi)$, and $s(\chi)$ without changing the scaling structure itself.



\end{document}